\documentclass[11pt]{article}

\usepackage[preprint]{acl}

\usepackage{times}
\usepackage{latexsym}
\usepackage{booktabs}
\usepackage{multirow}
\usepackage{amsmath}
\usepackage{amssymb}   
\usepackage[linesnumbered,ruled,vlined]{algorithm2e}
\usepackage{tcolorbox}
\usepackage{listings}
\tcbuselibrary{breakable, skins}
\usepackage{caption}
\usepackage{enumitem}
\usepackage{pifont}
\usepackage{xcolor}

\usepackage[T1]{fontenc}
\usepackage[utf8]{inputenc}

\usepackage{microtype}

\usepackage{inconsolata}

\usepackage{graphicx}

\title{World Craft: Agentic Framework to Create Visualizable Worlds via Text}

\newcommand{\eg}{\emph{e.g.}}

\newcommand{\vs}{\emph{vs.}}

\newcommand{\ws}{World Scaffold\space}
\newcommand{\wg}{World Guild\space}

\author{
Jianwen Sun $^{1,2,3*}$\quad 
Yukang Feng $^{1,2*}$\quad 
Kaining Ying $^{4*}$\quad 
Chuanhao Li $^{7}$\\
\textbf{Zizhen Li} $^{1,2,3}$ \quad
\textbf{Fanrui Zhang} $^{2}$ \quad
\textbf{Jiaxin Ai} $^{4,2}$\quad 
\textbf{Yifan Chang} $^{2}$\\ 
\textbf{Yu Dai} $^{3}$ \quad
\textbf{Yifei Huang} $^{1}$\quad 
\textbf{Kaipeng Zhang} $^{1,2\dagger}$ \\
[2mm]
$^1$ Shanda AI Research \quad
$^2$ Shanghai Innovation Institute \quad
$^3$ Nankai University \\
$^4$ Fudan University \quad
$^5$ Wuhan University \quad
$^6$ Shanghai AI Laboratory \\
[1mm]
\texttt{jianwen.sun@shanda.com, kaipeng.zhang@shanda.com$^{\dagger}$} \\
Project Page: https://github.com/HerzogFL/World-Craft
}

\begin{document}
\maketitle

\begin{abstract}
Large Language Models (LLMs) motivate generative agent simulation (\eg, AI Town) to create a ``dynamic world'', holding immense value across entertainment and research. However, for non-experts, especially those without programming skills, it isn't easy to customize a visualizable environment by themselves. In this paper, we introduce \textbf{World Craft}, an agentic world creation framework to create an executable and visualizable AI Town via user textual descriptions. It consists of two main modules, \textbf{\ws}and \textbf{World Guild}. 
\ws is a structured and concise standardization to develop interactive game scenes, serving as an efficient scaffolding for LLMs to customize an executable AI Town-like environment.
\wg is a multi-agent framework to progressively analyze users' intents from rough descriptions, and synthesizes required structured contents (\eg environment layout and assets) for \ws.
Moreover, we construct a high-quality error-correction dataset via reverse engineering to enhance spatial knowledge and improve the stability and controllability of layout generation, while reporting multi-dimensional evaluation metrics for further analysis. 
Extensive experiments demonstrate that our framework significantly outperforms existing commercial code agents (Cursor and Antigravity) and LLMs (Qwen3 and Gemini-3-Pro). in scene construction and narrative intent conveyance, providing a scalable solution for the democratization of environment creation. 

\end{abstract}

\vspace{-4mm}
\section{Introduction}
\vspace{-1mm}

\begin{figure}[t]
    \centering
    \includegraphics[width=0.98\linewidth]{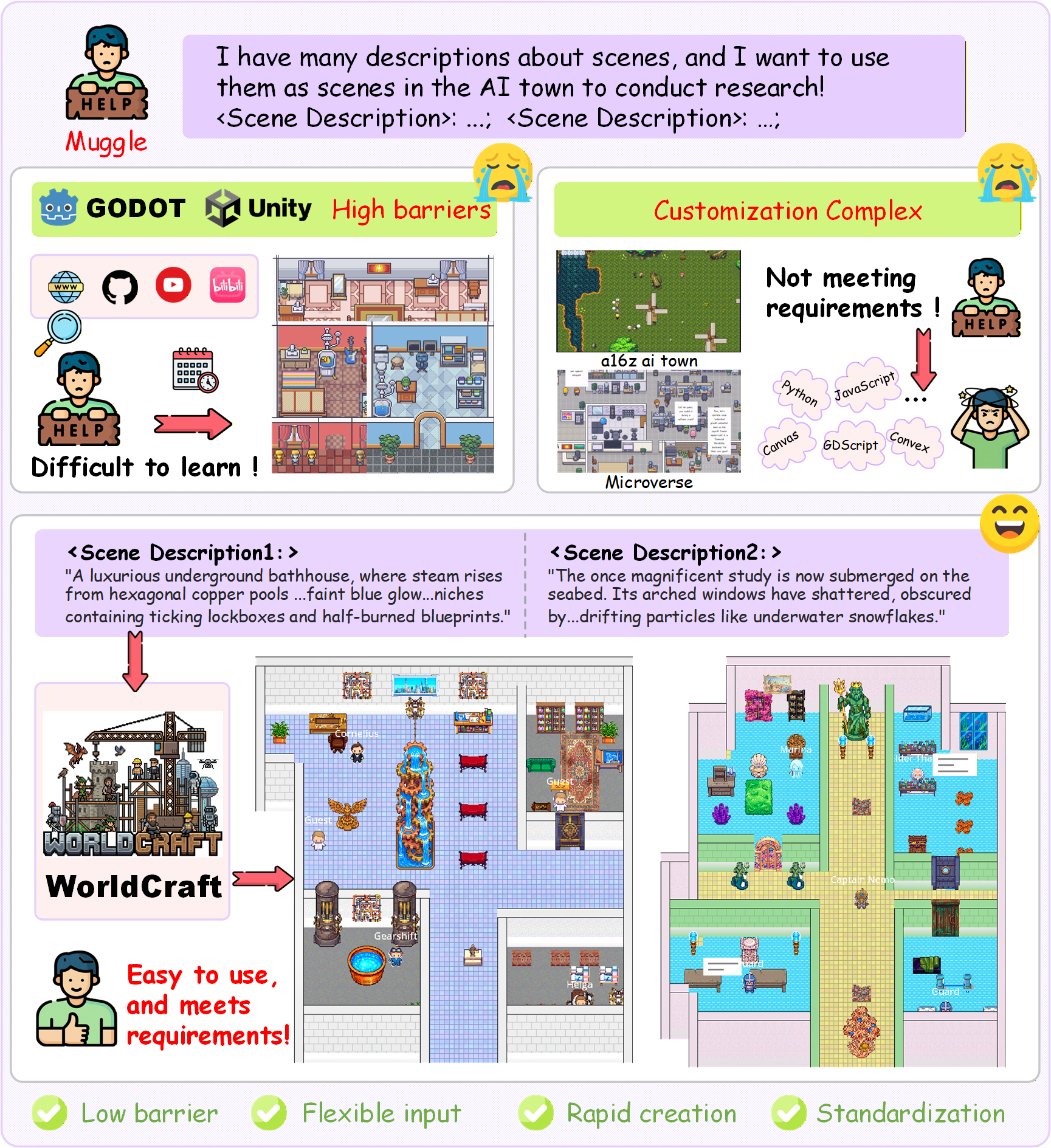}
    \vspace{-2mm}
    \caption{
        An illustration of our motivation and goal. 
        }
    \vspace{-6mm}
    \label{fig:teaser}
\end{figure}

AI Town represents a novel form of entertainment \cite{4_gong-etal-2024-mindagent,5_10.5555/3666122.3668483} and social simulation~\cite{1_yao2023react,2_wang-etal-2023-humanoid,3_xi2023risepotentiallargelanguage}, providing an ideal environment for observing complex emergent behaviors of agents. However, the construction of such environments still faces obstacles~\cite{6_li2024behavior1khumancenteredembodiedai}. Existing development workflows often rely on preset maps, suffer from fragmented toolchains and lack unified standards. They typically require users to possess professional programming skills (\eg, Unity or Godot)~\cite{7_10657821,59_lin2023agentsims}, posing a high barrier for those without programming backgrounds, limiting widespread adoption~\cite{8_xie2023openagentsopenplatformlanguage}. 
To address this, we face two main challenges: 
(1) The toolchains of traditional game engines are highly fragmented and complex to operate, lacking unified interfaces. This makes it difficult for AI agents to directly invoke low-level APIs for environment creation. 
(2) Human language is highly ambiguous. It is extremely challenging to directly model the precise layout content required for game environment construction from vague text descriptions.
To this end, we introduce \textbf{World Craft} as in Fig.1, comprising two subsystems: \textbf{\ws} and \textbf{\wg}. First, \ws serves as the infrastructure that automatically constructs executable game scenes from structured content, thereby accelerating the creation process and significantly lowering the entry barrier.

Although \ws bridges the underlying protocol, achieving fully automated construction requires relying on general LLMs to drive this process. However, addressing the second challenge encounters a fundamental obstacle: there is a significant semantic gap between abstract human narrative intents and the precise spatial instructions required for environment creation~\cite{9_Tan2018Text2SceneGC, 10_NEURIPS2021_64986d86, 11_10.5555/3666122.3666924}. Lacking embodied perception and precise spatial layout capabilities~\cite{12_10.5555/3666122.3669442,13_bisk-etal-2020-experience}, general LLMs often yield designs plagued by ``physical hallucinations'' such as floating objects or blocked paths~\cite{14_Tang_2023_ICCV,32_wu2023autogenenablingnextgenllm}. Inspired by research on Chain-of-Thought and modular reasoning~\cite{15_LI2025130919,16_10.5555/3666122.3667779}, we propose the \wg multi-agent framework. To mitigate the impact of the semantic gap, it decouples intent analysis from spatial planning, transforms complex cross-modal generation into a controllable step-by-step reasoning process, effectively improves the performance of LLMs, and leverages our built asset library to ensure the visual and physical consistency of the final output.

While \wg mitigates the impact of the semantic gap, general LLMs still face bottlenecks under complex geometric constraints due to a lack of spatial commonsense~\cite{17_stogiannidis2025mindgapbenchmarkingspatial,18_Jia2024SceneVerseS3}. Addressing the scarcity of high-quality layout data~\cite{19_Brazil2022Omni3DAL}, we utilize a ``Reverse Synthesis'' data construction paradigm. Instead of relying on expensive fully manual annotation~\cite{20_10.5555/3666122.3668522}, this method leverages ``Golden Layouts'' constructed via procedural algorithms, model verification, and minimal human correction. By applying reverse semantic restoration and controlled ``intentional corruption,'' it synthesizes full-chain supervision signals covering ``semantic mapping,'' ``generation from scratch,'' and ``error correction,'' thereby injecting key spatial reasoning and correction capabilities into the model~\cite{21_10.5555/3666122.3666499,22_ICLR2024_fef12656}. Combined with our proposed multi-dimensional evaluation benchmark~\cite{23_liu2025worldcraftphotorealistic3dworld}, our method demonstrates superior performance in logical correctness and intent conveyance.

\noindent Our contributions are summarized as follows:
\vspace{-2mm}

\begin{itemize}[leftmargin=10pt]
\setlength\itemsep{0em}
\item We propose \textbf{World Craft}, a framework integrating \textbf{\ws}and \textbf{\wg}which enables creating an interactive AI Town-like environments from natural language.
\item \textbf{\ws} is a flexible and standardized scaffold for LLMs to customize game environments. \textbf{\wg} mitigates the semantic gap from users' rough description to structured world and synthesize assets through multi-agent collaboration with step-by-step reasoning and our introduced high-quality asset library.%, while ensuring visual and physical consistency via an asset library.
\item We establish multi-dimensional evaluation metrics and construct a high-quality dataset which can effectively fill the knowledge gap of LLMs in complex spatial reasoning. Extensive experiments demonstrate that our framework significantly outperforms existing commercial code agents and LLMs.
\end{itemize}

\section{Related Works}
\vspace{-1mm}

\paragraph{Generative Agents.} Represented by Generative Agents ~\cite{24_10.1145/3586183.3606763}, the ``AI Town'' research initiated a wave of behavioral simulation. Subsequent works like Concordia ~\cite{25_mao2025agentkernelmicrokernelmultiagentframework}, AgentVerse ~\cite{26_ICLR2024_578e65cd}, and CAMEL ~\cite{27_10.5555/3666122.3668386} expanded the scope to group evolution. However, compared to advancements in agent memory and planning mechanisms, environment construction remains lagging: existing methods mostly rely on unmodifiable pre-built maps (\eg, Minecraft ~\cite{28_wang2023voyageropenendedembodiedagent,29_zhu2023ghostminecraftgenerallycapable}, 2D grids ~\cite{30_Park2022SocialSC}) or text-based sandboxes lacking physical properties ~\cite{33_ICLR2024_b3075b88}. Furthermore, existing open-source projects (Microverse, or TinyTroupe ~\cite{34_salem2025tinytroupellmpoweredmultiagentpersona})are built on different engines (\eg, Unity or Godot). This fragmentation of technical stacks significantly raises the barrier. Therefore, developing a standardized scenario construction tool is crucial for promoting the popularization of this field.

\begin{figure*}[t]
    \centering
    \includegraphics[width=\linewidth]{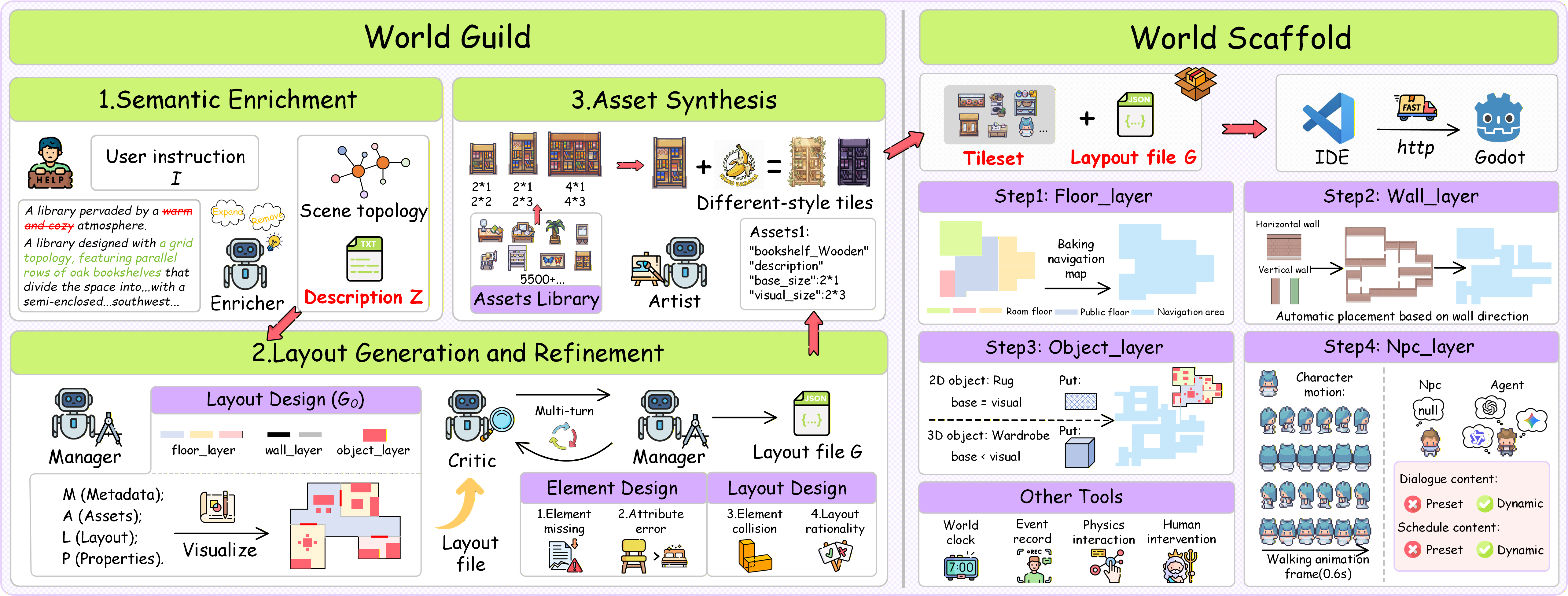}
    \vspace{-6mm}
    \caption{
        Architecture of WorldCraft. It comprises the World Guild for intent analysis and layout generation, and the World Scaffold for automated scene construction.
        }
    \vspace{-6mm}
    \label{fig:framework}
\end{figure*}

\paragraph{Layout Generation.} Works such as House-GAN++, HouseDiffusion, FloorPlan-LLaMa and others~\cite{35_9577959,36_Shabani2022HouseDiffusionVF,37_yin-etal-2025-floorplan, 57_leng-etal-2023-tell2design} have demonstrated capabilities in topological layout generation, while SceneCraft ~\cite{38_10.5555/3692070.3692846} and 3D-GPT ~\cite{39_11125634} explored text-to-3D visual synthesis. Unlike these studies that focus on visual or geometric aspects, this paper is dedicated to generating structured layouts with complete functional logic. However, as noted by Mind's Eye ~\cite{40_liu2022mindseyegroundedlanguage}, Spatial-VLM ~\cite{41_10658310}, and PlanQA ~\cite{42_rodionov2025floorplanqabenchmarkspatialreasoning}, general LLMs lack embodied perception and suffer from a ``semantic gap'' when mapping abstract language to physical constraints. Consequently, end-to-end generation models struggle to ensure the correctness of spatial logic. To address this, we introduce a multi-agent collaboration mechanism to decouple intent parsing from spatial planning, significantly reducing generation difficulty through stepwise reasoning.

\paragraph{Knowledge Enhancement.} LLMs still suffer from a knowledge deficit when handling complex spatial layouts and geometric constraint~\cite{55_Sun2024LayoutVLMDO}. To address this lack of domain knowledge, mainstream methods employ RAG~\cite{43_NEURIPS2020_6b493230,44_10.5555/3524938.3525306} or utilize instruction fine-tuning~\cite{45_NEURIPS2022_b1efde53,46_wang-etal-2023-self-instruct} to align task distributions. However, in the layout domain, existing open-source datasets primarily focus on static visual representations~\cite{47_10205395,48_Fu20203DFRONT3F}, lacking high-quality ``instruction-layout'' pairs~\cite{49_NEURIPS2023_413885e7}. Furthermore, studies~\cite{50_NEURIPS2023_91edff07,51_ICLR2024_2460396f} indicate that iterative correction capability is equally critical for resolving complex constraints.%, yet such data is almost non-existent. 
To address this data scarcity, we utilize a reverse data construction method to generate a large-scale corpus containing correction trajectories, and adopt a two-stage training strategy to equip the model with professional layout planning and self-correction capabilities~\cite{52_NEURIPS2022_9d560961}.

\vspace{-1mm}
\section{Method}
\vspace{-1mm}
\subsection{Problem Formulation}
\vspace{-1mm}

We define text-based game scene design as a mapping from natural language instruction $\mathcal{I}$ to structured layout $\mathcal{G}$. It is formalized as a quadruple (see Appendix~\ref{sec:appG} for examples):
\begin{equation}
    \mathcal{G} = (M, A, L, P)
\end{equation}
where $M$ (Metadata) defines the overall scene style and grid size; $A$ (Assets) describes the visual style and layer attributes; $L$ (Layout) records the precise spatial coordinates of components; and $P$ (Properties) specifies interaction properties.
Given that $\mathcal{I}$ typically implies ambiguous narrative intents while $\mathcal{G}$ demands determinate geometric parameters and physical properties, directly modeling $P(\mathcal{G} | \mathcal{I})$ faces a significant semantic gap. For this, we introduce an intermediate variable $\mathcal{Z}$ as a semantic bridge-representing scene topology and functional distribution without specific coordinates. By decomposing the generation objective into:
\begin{equation}
    P(\mathcal{G} | \mathcal{I}) = \sum_{\mathcal{Z}} P(\mathcal{Z} | \mathcal{I}) \cdot P(\mathcal{G} | \mathcal{Z}),
\end{equation}
we achieve a logical decoupling of parsing intents before grounding parameters.
\vspace{-1mm}
\subsection{Collaborative Multi-Agent Framework}

To effectively solve the aforementioned decomposition process, we propose the \wg framework. This framework introduces a multi-agent collaboration mechanism designed to alleviate the semantic gap inherent in direct modeling by decoupling intent parsing from spatial planning. Through step-by-step reasoning, it transforms the direct cross-modal mapping into a series of executable logical operations, thereby significantly reducing generation difficulty. \wg consists of four core agents: Semantic Enrichment (Enricher), Layout Generation (Manager), Quality Assurance (Critic), and Asset Synthesis (Artist). As shown in Fig.~\ref{fig:framework}, the detailed functional descriptions of each agent are as follows.

\paragraph{Semantic Enrichment}

The Enricher is responsible for transforming the user instruction $\mathcal{I}$ into a layout description $\mathcal{Z}$ endowed with spatial logic. Since user inputs often exhibit significant disparities in information density-ranging from sparse keywords to abstract descriptions that are difficult to ground, the Enricher needs to concretize narrative intents into a coherent scene topology, explicitly defining connectivity and the rough distribution of core components. This process does not involve specific coordinate calculations but focuses on constructing a spatial sketch that is consistent with common sense and logically self-consistent, thereby eliminating ambiguity at the semantic level and providing guidance for the subsequent precise design by the Manager.
\vspace{-1mm}
\paragraph{Constrained Layout Generation}

The Manager is responsible for executing the grounding process of $P(\mathcal{G} | \mathcal{Z})$. It receives the layout description $\mathcal{Z}$ from the Enricher and converts it into an initial layout file $\mathcal{G}_0$ that conforms to physical definitions. As the core planning agent, the Manager's function is to parse the topological logic and relative positional constraints contained in the natural language and map them into quantitative, precise geometric parameters. Specifically, guided by $\mathcal{Z}$, it determines the scene metadata $M$, instantiates the asset library $A$ and property set $P$, and designs the grid coordinates and orientation for each component in the layout layer $L$. This outputs an initial layout file with a complete hierarchy and asset attributes, achieving the cross-modal transformation from abstract text descriptions to executable data.

\begin{figure*}[t]
    \centering
    \includegraphics[width=\linewidth]{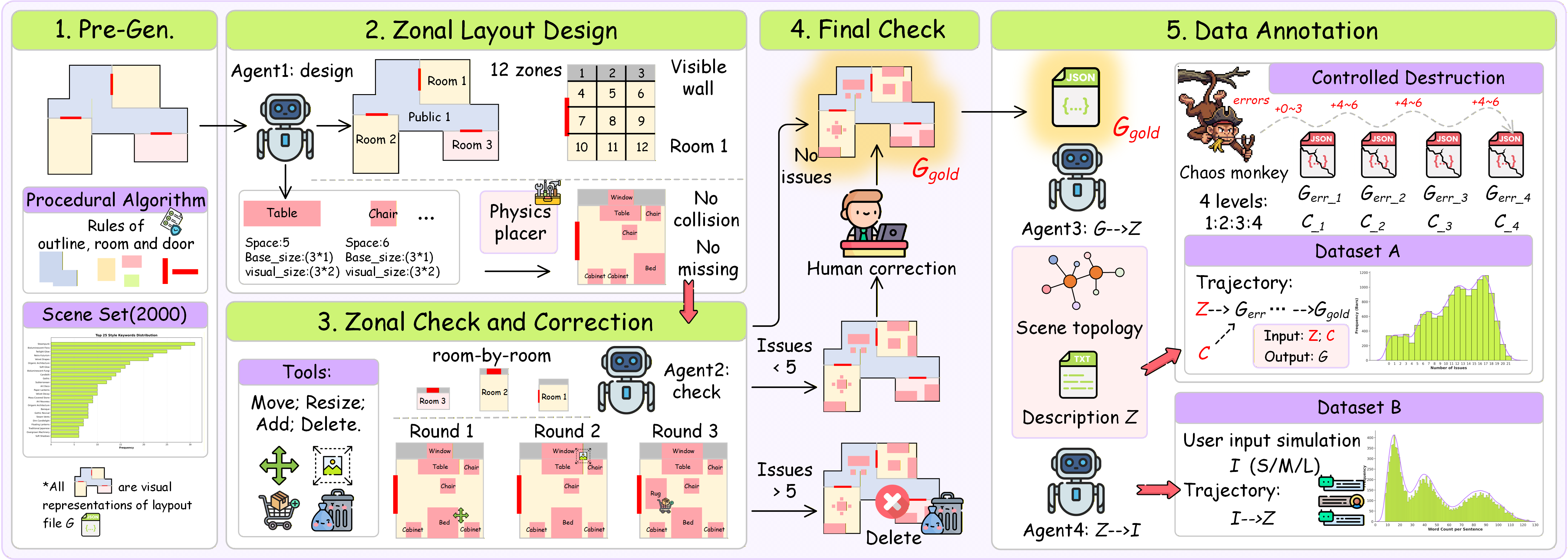}
        \vspace{-6mm}
    \caption{
        Two-stage fine-tuning data construction process. Utilizing Gemini-3-Pro as all the agents, we perform 10 runs for each of scenario descriptions. During the filtering process, approximately 5k invalid samples are discarded, and 1.2k long-tail samples undergo human rectification, resulting in a final dataset of approximately 14k samples.
        }
    \vspace{-6mm}
    \label{fig:pipeline}
\end{figure*}
\vspace{-1mm}
\paragraph{Iterative Critique and Refinement}

To ensure the generated results meet physical and logical constraints, we introduce the Critic to establish an iterative feedback loop. In the $t$-th iteration, the Critic performs rule-based physical checks (such as collision and connectivity detection) and model-based semantic evaluations on the current layout $\mathcal{G}_t$, generating specific correction instructions $\mathcal{C}_t$. If defects are detected (i.e., $\mathcal{C}_t \neq \emptyset$), the Manager executes targeted spatial editing operations (such as moving object coordinates or replacing assets) based on these instructions to generate a corrected layout $\mathcal{G}_{t+1}$. This process continues until all checks are passed or the maximum number of rounds $T_{max}$ is reached, ensuring the rationality and logical self-consistency of the final output layout.
\vspace{-1mm}
\paragraph{Reference-Guided Asset Synthesis}

The Artist is responsible for transforming the asset definition set $A$ within the layout design $\mathcal{G}$ into visual assets. To address the common issue of style fragmentation in pure text-to-image generation, we employ a retrieval-augmented texture synthesis strategy: for each component, the Artist first retrieves a reference image $v_{ref}$ from the pre-built library $\mathcal{D}_{lib}$ (see Appendix~\ref{sec:appC} for library examples and algorithm details). Using this as a style anchor to guide the generative model, it produces Tile resources that possess a unified visual style while maintaining semantic accuracy. Finally, the \ws automatically assembles the generated visual resources with the layout layer $L$ and property set $P$, constructing a playable game scene complete with navigation meshes and interaction logic.

\subsection{Data Construction}

Although \wg mitigates the impact of the semantic gap, LLMs still face performance bottlenecks under complex geometric constraints due to a lack of spatial commonsense~\cite{53_NEURIPS2024_89cc5e61,54_xu2025origamispacebenchmarkingmultimodalllms,60_10.1007/978-3-031-73337-6_9}. To equip LLMs with professional layout planning and logical correction capabilities, we designed a data construction pipeline comprising three stages: first, Scenario Initialization establishes the diversity of scene configurations; then, Scene Design combines procedural rules and verification to construct the golden layout $\mathcal{G}_{gold}$; finally, Data Annotation applies controlled degradation to them to generate fine-tuning data with complete correction trajectories.
\vspace{-1mm}
\paragraph{Scenario Initialization}
\label{sec:data}

To ensure data coverage and generalization, we constructed a base scenario library spanning four dimensions: real-world, literature, film, and games. We selected 125 seed scenarios per category, partitioned into training and held-out test sets with a 4:1 ratio to prevent data leakage. For the training set, we established a prompt pool containing 560 style descriptions and randomly injected 5 variants (\eg, ``Cyberpunk'', ``Primitive'') into each scenario, expanding the dataset to 2,000 samples. This strategy aims to enhance the model's spatial logic robustness across cross-domain scenarios by leveraging highly diverse semantic atmospheres. Details of the scenario data are provided in Fig.~\ref{fig:pipeline} and the Appendix~\ref{sec:appF}.
\vspace{-1mm}
\paragraph{Scene Design}

To construct the golden layout $\mathcal{G}_{gold}$ satisfying strict physical constraints, we designed an offline generation pipeline with multi-stage verification. First, procedural algorithms(see Appendix~\ref{sec:appH}) generate empty room structures; then, LLM assigns specific functional attributes based on scene descriptions. During the filling stage, to overcome the spatial perception deficits of LLMs, we introduced the ``12-zone grid'' strategy to assist relative orientation generation. This strategy partitions each room into left-center-right visible walls and an internal 9-grid, guiding the model to output component coordinates based on relative orientations, while coordinating with a ``Physical Placer'' to eliminate collision conflicts in real-time. Finally, a Teacher Model equipped with editing tools is employed to automatically review and refine the layout room-by-room, supplemented by human experts for long-tail samples, thereby ensuring the logical and physical rigor of all $\mathcal{G}_{gold}$ data.

\paragraph{Data Annotation}

Based on the constructed golden layouts $\mathcal{G}_{gold}$, we first utilize a LLM to reverse-engineer them into coordinate-free layout descriptions $\mathcal{Z}$, serving as a unified semantic foundation. Subsequently, we introduce a ``Chaos Monkey'' (perturbation agent) to execute four levels of controlled destruction with weights of 1:2:3:4 (\eg, components exchange or creating collisions), generating error samples containing 2 to 15 issues along with correction instructions: $(\Phi(\mathcal{G}_{gold}) \rightarrow\mathcal{G}_{error}, \mathcal{C})$. On this basis, we define two core datasets. First, we construct \textbf{Dataset A}:
\begin{equation}
    \mathcal{D}_{A} = \{ \mathcal{Z} \!\to\! \mathcal{G}_{gold} \} \cup \{ (\mathcal{G}_{err}, \mathcal{C}) \!\to\! \mathcal{G}_{gold} \},
\end{equation}
which records the trajectory from generating initial layouts via $\mathcal{Z}$ to iteratively repairing $\mathcal{G}_{error}$ into $\mathcal{G}_{gold}$ using $\mathcal{C}$. 
% This sequence data enables the Manager to both translate descriptions into designs and execute error-fixing instructions.
Secondly, we constructed \textbf{Dataset B} by simulating users rewriting $\mathcal{Z}$ into natural language instructions $\mathcal{I}$ of three densities (short, medium, and long), forming paired data mapping user inputs to layout descriptions dedicated:
% to training the Enricher's semantic parsing capabilit
\begin{equation}
    \mathcal{D}_{B} = \{ (\mathcal{I}, \mathcal{Z}) \mid \mathcal{I} \sim \text{Sim}_{user}(\mathcal{Z}, \rho) \},
\end{equation}
where $\rho \in \{\text{short}, \text{medium}, \text{long}\}$ denotes the simulated instruction density. Fig.~\ref{fig:pipeline} illustrate the complete construction flow and data distribution details.

\subsection{Training Strategy}

Based on the aforementioned high-quality datasets, we adopt a decoupled two-stage fine-tuning strategy to specifically optimize semantic understanding and spatial execution capabilities.

\paragraph{Semantic Alignment.} The first stage aims to endow the Enricher (parameterized by $\theta_E$) with intent normalization capabilities. Utilizing dataset $\mathcal{D}_B$, we establish a deterministic mapping from arbitrary natural language $\mathcal{I}$ to standard layout descriptions $\mathcal{Z}$ by maximizing the conditional likelihood of semantic tokens:
\begin{equation}
\begin{split}
    \mathcal{L}_{\text{SFT}}^{(E)}(\theta_E) = & -\mathbb{E}_{(\mathcal{I}, \mathcal{Z}) \sim \mathcal{D}_B} \\
    & \sum_{t=1}^{|\mathcal{Z}|} \log P_{\theta_E}(z_t \mid \mathcal{I}, z_{<t}).
\end{split}
\end{equation}

Under this objective, by mixing instruction data of varying densities, the model acquires robust normalization capabilities: it can perform commonsense logical completion for sparse instructions while extracting key topological information from verbose descriptions.

\paragraph{Spatial Refinement.} The second stage is focusing on enhancing the Manager's (parameterized by $\theta_M$) spatial planning and dynamic correction capabilities. Based on dataset $\mathcal{D}_A$, we unify the tasks of ``initial generation from $\mathcal{Z}$'' and ``correction based on $\mathcal{C}$'' into a sequence prediction format. Defining the input context as $\mathcal{X} \in \{\mathcal{Z}, (\mathcal{G}_{error}, \mathcal{C})\}$, the optimization objective is:
\begin{equation}
\begin{split}
    \mathcal{L}_{\text{SFT}}^{(M)}(\theta_M) = & -\mathbb{E}_{(\mathcal{X}, \mathcal{G}_{\text{gold}}) \sim \mathcal{D}_A} \\
    & \sum_{t=1}^{|\mathcal{G}|} \log P_{\theta_M}(g_t \mid \mathcal{X}, g_{<t}).
\end{split}
\end{equation}
This strategy not only enables the model to master the logic of converting layout descriptions into quadruplets $\mathcal{G}$ but also endows it to respond to correction instructions $\mathcal{C}$. Consequently, the model can execute precise editing operations upon receiving negative feedback from the Critic to optimize error states. Specific training settings and hyperparameter configurations are detailed in the Experiments section and Appendix~\ref{sec:appB}.

\section{Experiments}

\begin{table*}[t]
\centering
\setlength{\tabcolsep}{4pt}
\resizebox{\textwidth}{!}{%
\begin{tabular}{llcccccccc}
\toprule
& \textbf{Methods} & \textbf{CFR} $\uparrow$ & \textbf{RCS} $\uparrow$ & \textbf{OPS} $\downarrow$ & \textbf{CER} $\uparrow$ & \textbf{OVD} $\uparrow$ & \textbf{PAC} $\downarrow$ & \textbf{VSA-C} $\uparrow$ & \textbf{VSA-V} $\uparrow$ \\
\midrule
\multicolumn{10}{c}{\textit{Part 1: Effectiveness of the Inference Framework Design}} \\
\midrule
\multirow{3}{*}{\shortstack[l]{Direct Gen.\\(Few-shot)}} 
 & Base: Qwen3-32B & 0.59 & 0.44 & 9.27 & 0.76 & 3.27 & 10.55 & 20.22 & 3.73 \\
 & Open: Qwen3-235B & 0.73 & 0.60 & 7.18 & 0.80 & 4.09 & 7.22 & 23.34 & 4.74 \\
 & Closed: Gemini-3-Pro & 0.80 & 0.59 & 5.56 & 0.86 & 5.02 & 6.87 & 23.29 & 4.67 \\
\midrule
\multirow{3}{*}{\shortstack[l]{Direct Gen.\\+ Critic}} 
 & Base: Qwen3-32B & 0.66 & 0.49 & 8.68 & 0.78 & 3.43 & 9.31 & 20.24 & 3.88 \\
 & Open: Qwen3-235B & 0.78 & 0.67 & 6.88 & 0.83 & 4.37 & 6.17 & 23.73 & 5.03 \\
 & Closed: Gemini-3-Pro & 0.84 & 0.67 & 4.99 & 0.90 & 5.30 & 5.92 & 23.51 & 4.98 \\
\midrule
\multirow{3}{*}{\shortstack[l]{Enricher\\+Manager\\+Critic}} 
 & Base: Qwen3-(8+32)B$^*$ & 0.66 & 0.52 & 7.47 & 0.77 & 4.46 & 9.33 & 21.45 & 4.03 \\
 & Open: Qwen3-235B$^*$ & 0.81 & 0.71 & 4.72 & 0.84 & 5.28 & 6.13 & 24.72 & 5.66 \\
 & Closed: Gemini-3-Pro $^*$ & 0.83 & 0.68 & 3.77 & 0.92 & 5.92 & 5.90 & 25.08 & 5.71 \\
\midrule
\multicolumn{10}{c}{\textit{Part 2: Effectiveness of Data and Training Strategies}} \\
\midrule
\multirow{6}{*}{\shortstack[l]{Training Variants\\(w/o Critic)}} 
 & End-to-End Fine-tuning & 0.86 & 0.78 & 5.36 & 0.87 & 6.12 & 4.81 & 24.83 & 5.33 \\
 & Qwen3-(8+8)B & 0.83 & 0.73 & 4.64 & 0.90 & 6.35 & 5.42 & 25.31 & 5.64 \\
 & Qwen3-(8+32)B & 0.88 & 0.79 & 3.93 & 0.92 & 7.10 & 4.76 & 26.36 & 6.17 \\
 & Stage 1 Only (Enricher) & 0.63 & 0.46 & 8.16 & 0.76 & 3.97 & 9.88 & 25.22 & 5.22 \\
 & Stage 2 Only (Manager) & 0.89 & 0.78 & 4.11 & 0.91 & 6.84 & 4.76 & 22.16 & 4.34 \\

\midrule
\multirow{2}{*}{\shortstack[l]{Two-Stage\\+ Critic}} 
 & w/ Standard Data$^\dagger$ & 0.89 & 0.81 & 3.67 & 0.94 & 7.07 & 4.39 & 26.83 & 6.04 \\
 & \textbf{w/ Correction Data(Ours)}$^* $$^\dagger$ & \textbf{0.94} & \textbf{0.88} & \textbf{3.03} & \textbf{0.99} & \textbf{7.13} & \textbf{3.64} & \textbf{28.07} & \textbf{6.80} \\
\bottomrule
\end{tabular}
}
\caption{\textbf{Experimental results of scene generation.} 
The top and bottom sections validate the framework design and the data/training strategies, respectively. The critic is based on GPT-5.1 with max rounds $T=4$. Note that for \textbf{VSA-C}, evaluation is strictly limited to samples with fewer than 77 tokens due to CLIP's length constraint.
}
\label{tab:main_results}
\end{table*}

\subsection{Experimental Setup}

\paragraph{Implementation Details and Baselines.}
Following the proposed two-stage training strategy, we fine-tuned the Qwen3 series (8/32B)~\cite{58_qwen3} open-source models, utilizing different model sizes to explore the optimal balance between performance and efficiency. Given the lack of specialized models for such structured spatial reasoning tasks, we focused the comparison on general-purpose LLMs: we selected Qwen3-235B as the performance upper bound of open-source models (Open SOTA) and Gemini-3-Pro to represent the peak level of closed-source commercial models (Closed SOTA), thereby establishing a widely representative and fair comparison baseline.

\paragraph{Evaluation Datasets and Metrics.}
\label{sec:metrics}

To ensure objective evaluation, we constructed a manually annotated test set derived from the scene library in Section~\ref{sec:data}. Specifically, we selected 100 ``held-out'' seeds (25 each from Reality, Literature, Film, and Games) strictly excluded from training. We employed an LLM to generate instructions of three complexity levels (Short, Medium, Long) for each seed, followed by expert refinement, yielding 300 test samples. We evaluate eight core metrics across three dimensions: Layout Design via Collision-Free Rate (CFR), Room Connectivity Score (RCS), and Object Placement Score (OPS); Element Design via Component Existence Rate (CER), Object Volume Density (OVD), and Property Consistency (PAC); and Intent Alignment using VSA-C (CLIP) and VSA-V (VLM) to verify visual-semantic consistency. The detailed prompt for each metric can be found in the Appendix~\ref{sec:appE}.

\begin{table*}[t]
\centering
\setlength{\tabcolsep}{4pt}
\resizebox{\textwidth}{!}{%
\begin{tabular}{lccccccccccc}
\toprule
& \multicolumn{4}{c}{\textbf{Dim 1: Layout Rationality}} & \multicolumn{4}{c}{\textbf{Dim 2: Element Richness}} & \multicolumn{3}{c}{\textbf{Dim 3: Visual Consistency}} \\
\cmidrule(lr){2-5} \cmidrule(lr){6-9} \cmidrule(lr){10-12}
\textbf{Method} & \textbf{CFR} $\uparrow$ & \textbf{RCS} $\uparrow$ & \textbf{OPS} $\downarrow$ & \textbf{HWR} $\uparrow$ & \textbf{CER} $\uparrow$ & \textbf{OVD} $\uparrow$ & \textbf{PAC} $\downarrow$ & \textbf{HWR} $\uparrow$ & \textbf{VSA-C} $\uparrow$ & \textbf{VSA-V} $\uparrow$ & \textbf{HWR} $\uparrow$ \\
\midrule
Base: Qwen3-(8+32)B & 0.66 & 0.52 & 7.47 & 0.28 & 0.77 & 4.46 & 9.33 & 0.19 & 21.45 & 4.03 & 0.38 \\
Open: Qwen3-235B & 0.81 & 0.71 & 4.72 & 0.47 & 0.84 & 5.28 & 6.13 & 0.45 & 24.72 & 5.66 & 0.46 \\
Closed: Gemini-3-Pro & 0.83 & 0.68 & 3.77 & 0.46 & 0.92 & 5.92 & 5.90 & 0.53 & 25.08 & 5.81 & 0.47 \\
Ours & 0.94 & 0.88 & 3.03 & 0.78 & 0.99 & 7.13 & 3.64 & 0.82 & 28.07 & 6.80 & 0.69 \\
\midrule
\textit{Reliability} & $r=0.95$ & $r=0.96$ & $r=-0.86$ & $\kappa=0.61$ & $r=0.97$ & $r=0.99$ & $r=-0.98$ & $\kappa=0.54$ & $r=0.98$ & $r=0.91$ & $\kappa=0.65$ \\
\bottomrule
\end{tabular}
}
\caption{\textbf{Human Evaluation and Metric Correlation.} Comparison between automated metrics and Human Win Rate (HWR) across three dimensions. The strong Pearson correlation ($|r|$) and substantial Fleiss' Kappa ($\kappa$) validate that our automated metrics are reliable proxies for human preference.}
\vspace{-4mm}
\label{tab:reliability_metrics}
\end{table*}

\begin{figure*}[t]
    \centering
    \includegraphics[width=\linewidth]{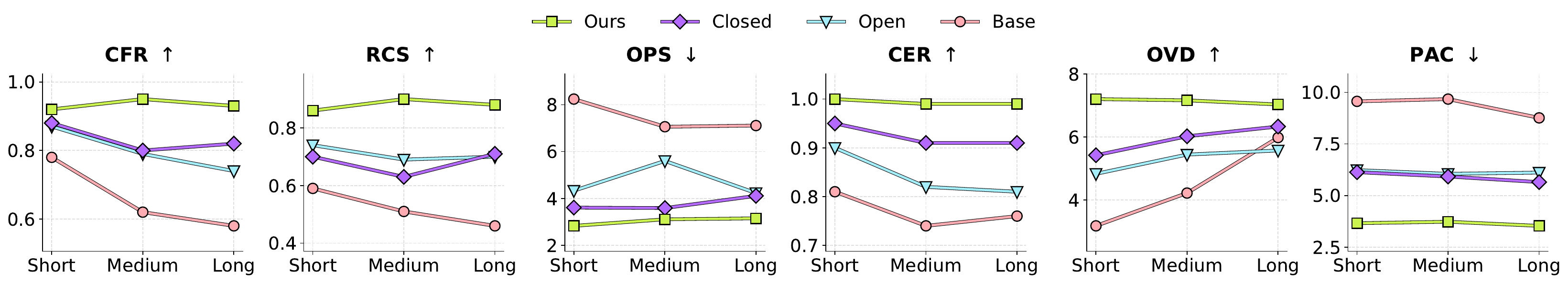}
    \vspace{-8mm}
    \caption{
        Results of fine-grained comparison on performance stability under different input lengths in the test set.
        }
    \vspace{-4mm}
    \label{fig:radar}
\end{figure*}

\begin{figure*}[t]
    \centering
    \includegraphics[width=\linewidth]{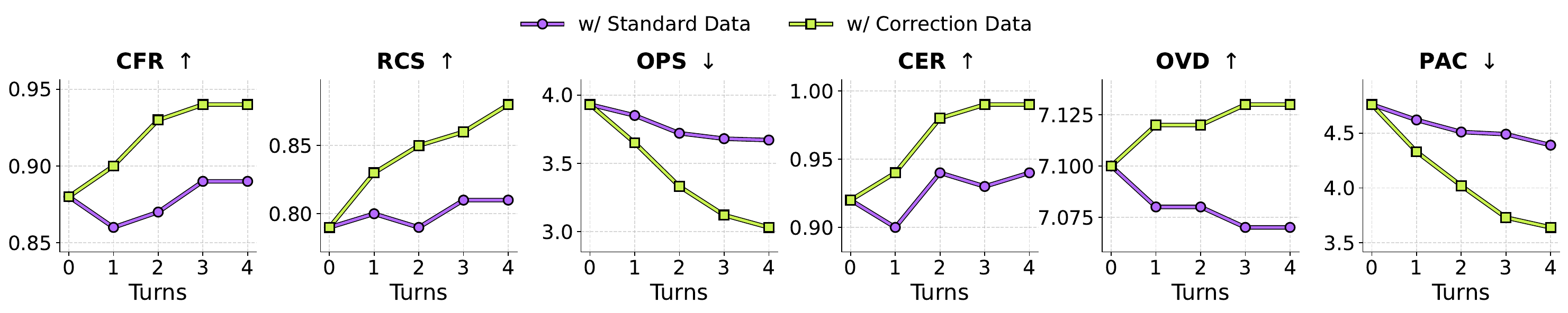}
    \vspace{-8mm}
    \caption{
        Dynamic changes in model output quality during multi-round correction processes.
        }
    \vspace{-4mm}
    \label{fig:line}
\end{figure*}

\subsection{Main Results}
\paragraph{Framework Design.} The upper part of Table \ref{tab:main_results} validates the effectiveness of the stepwise reasoning design. First, the necessity of the Critic module is confirmed: introducing the Critic to Direct Gen. baseline yields significant improvements in layout design. Second, the effectiveness of stepwise reasoning (Enricher+Manager) is further demonstrated: with the introduction of stepwise reasoning, the model achieves notable gains in metrics such as RCS, OPS, and OVD. This indicates that decoupling the complex generation task into two sub-steps-``semantic completion'' and ``spatial management''-allows each module to focus on specific tasks, thereby verifying the rationality of the framework design.

\paragraph{Data and Training Strategy.} The lower part of Table~\ref{tab:main_results} validates the rationality of our training strategy. First, decoupled training outperforms end-to-end fine-tuning, demonstrating the necessity of separating semantic and spatial tasks. Second, the (8+32)B combination outperforms the (8+8)B version in multiple aspects, indicating that the spatial planning task demands higher model capacity. Finally, we observe that when trained solely on standard data, the model possesses generation capabilities but struggles to make correct revisions based on feedback. In contrast, models trained on correction data benefit significantly during the correction phase. This proves the effectiveness of ``error-correction''.% of positive role of introducing ``error-correction'' trajectories during training.

\subsection{Fine-grained Analysis}
\paragraph{Instruction Robustness.}

To analyze model performance under different input conditions, we conducted a fine-grained breakdown of the data in Table \ref{tab:main_results}. Fig.~\ref{fig:radar} records the results for the four models marked with ``*'' across three instruction lengths. The data shows that general models exhibit significant performance fluctuations across different instruction lengths, indicating a struggle to cope with varying information densities and generate stable results. In contrast, our method maintains stability and infers reasonable layouts. This proves that our training strategy successfully establishes a mapping from abstract instructions to layout descriptions, granting the model robustness in handling instruction ambiguity.

\paragraph{Correction Trajectory.}

To explore the dynamic process of iterative critique and refinement, we conducted a fine-grained breakdown of the data in Table \ref{tab:main_results}. Fig.~\ref{fig:line} records the result changes across correction rounds $T=0, 1, \dots, 4$ for the two models marked with $\dagger$ in the table. Results: The model trained using only standard data, despite decent initial performance, shows a flat metric improvement during the multi-round correction process. In contrast, the model trained on correction data exhibits a robust growth trend, particularly in spatial layout metrics. This indicates that correction data is crucial for the model to correctly understand and execute modification instructions, ensuring the effectiveness of iterative optimization.

\subsection{Human Evaluation and Metric Validation}
\label{sec:humaneval}
To verify whether the metrics accurately reflect human perception of generation quality, we organized a subjective evaluation with 5 game players. The experiment adopted a pairwise forced-choice format, covering the four models marked with ``*'' in Table \ref{tab:main_results}, with 150 instructions randomly sampled from the test set (50 each for short, medium, and long). We summarized the eight metrics into three questions for the evaluators.(The content of the questions and details of the reviewers are provided in the Appendix~\ref{sec:appD}).We used the Pearson coefficient $|r|$ to report the consistency between metrics and human evaluation, and Fleiss' $\kappa$ to report Inter-Annotator Agreement. As shown in Table \ref{tab:reliability_metrics}, the results demonstrate a strong correlation between our metrics and human preference (mean Pearson's $|r|$>0.90), and evaluators reached substantial agreement (mean $\kappa$ = 0.60). Specifically, the highest consensus was reached on Visual Consistency ($\kappa$ = 0.65), while slight divergence was observed in Element Richness ($\kappa$ = 0.54). Overall, the results robustly validate the scientific validity and reliability of the proposed automated evaluation system.

\subsection{Comparison with Code Agents}
\label{sec:codeganet}
To verify the superiority of our framework (\wg) in creating generative agent simulation environments, we compared it with general code agents (Cursor and Antigravity) using the same Gemini-3-Pro. Setup: Three operators tested 15 prompts of different lengths (5 short, 5 medium, 5 long). General agents allowed multi-turn human debugging (max 60 mins), recording Time-to-Runnable (TR) and Time-to-Satisfaction (TS) in minutes; our method used fully automated one-shot generation. Evaluation: Five evaluators and a VLM(in Appendix~\ref{sec:appE}) performed double-blind pairwise comparisons. The results are shown in Table \ref{tab:comparison_agents}: In terms of efficiency, our method has a significant advantage in construction speed. In terms of quality, despite human corrections, our method still achieved the highest win rates in the evaluation. This proves that our method not only lowers technical barriers but also constructs high-fidelity generative agent simulation environments with exceptional speed.

\begin{table}[t]
\centering
\setlength{\tabcolsep}{4pt}
\resizebox{\columnwidth}{!}{%
\begin{tabular}{lcccc}
\toprule
\textbf{Methods} & \textbf{TR (min)} $\downarrow$ & \textbf{TS (min)} $\downarrow$ & \textbf{HWR} $\uparrow$ & \textbf{VWR} $\uparrow$ \\
\midrule
Cursor       & 15.42 & 48.15 & 0.23 &    0.35 \\
Antigravity  & 9.83 & 42.40 & 0.35 & 0.39\\
\textbf{Ours} & \textbf{4.25} & \textbf{4.25} & \textbf{0.92} & \textbf{0.76} \\
\bottomrule
\end{tabular}
}
\caption{\textbf{Comparison with Code Agents.} HWR/VWR: Human/Visual evaluation win rates. Tile generation: $\sim$20s/image (Nano Banana Pro, 8 threads).}
\label{tab:comparison_agents}
\vspace{-6mm}
\end{table}

\subsection{Ablation Study on Visual Generation}
To verify the role of the asset library in unifying visual styles, we performed an ablation comparison on the four models marked with ``*'' in Table~\ref{tab:main_results} (see Table~\ref{tab:ablation_library}). We use VGG Gram matrix distance to measure style differences and Visual Harmony (VH) to assess visual perception. Results show that removing the asset library leads to a sharp increase in VGG Loss and a significant drop in VH for all models, demonstrating that unprocessed tiles suffer from severe style discrepancies. Furthermore, comparing evaluation metrics reveals that VSA-C remains stable, while VSA-V shows a slight decline without the asset library. This indicates that inconsistent art styles interfere with the VLM's judgment. In summary, the asset library effectively resolves style discrepancy issues and ensures the visual consistency of generated scenes.

\begin{table}[t]
\centering
\setlength{\tabcolsep}{4pt}
\resizebox{\columnwidth}{!}{%
\begin{tabular}{lcccc}
\toprule
\textbf{Methods} & \textbf{VGG Loss} $\downarrow$ & \textbf{VH} $\uparrow$ & \textbf{VSA-C} $\uparrow$ & \textbf{VSA-V} $\uparrow$ \\
\midrule
\multicolumn{5}{c}{\textit{Part 1: Generation w/o Asset Library}} \\
\midrule
Base: Qwen3-(8+32)B & 259.03 & 4.62 & 21.67 & 4.21 \\
Open: Qwen3-235B & 273.19 & 5.09 & 25.02 & 5.17 \\
Closed: Gemini-3-Pro & 229.33 & 4.62 & 25.07 & 5.20 \\
\textbf{Ours} & 248.62 & 4.47 & 27.93 & 6.35 \\
\midrule
\multicolumn{5}{c}{\textit{Part 2: Generation w/ Asset Library}} \\
\midrule
Base: Qwen3-(8+32)B & 56.77 & 7.56 & 21.45 & 4.03 \\
Open: Qwen3-235B & 58.62 & 7.17 & 24.72 & 5.66 \\
Closed: Gemini-3-Pro & 56.63 & \textbf{7.62} & 25.08 & 5.71 \\
\textbf{Ours} & \textbf{55.80} & 7.36 & \textbf{28.07} & \textbf{6.80} \\
\bottomrule
\end{tabular}
}
\caption{\textbf{Visual Generation Ablation.} Top/Bottom: w/o \vs w/ asset library. VH uses flattened tiles to exclude layout influence. Prompts in the Appendix~\ref{sec:appE}.}
\label{tab:ablation_library}
\vspace{-5mm}
\end{table}

\vspace{0.5mm}
\section{Conclusion}
\vspace{0.5mm}
We present \textbf{World Craft}, integrating \textbf{\ws}, a standardized infrastructure addressing the issues of fragmented toolchains and high technical barriers, and \textbf{\wg}, a multi-agent framework mitigating the semantic gap between narrative and spatial instructions. Additionally, we introduce a ``reverse synthesis'' method to generate high-quality supervision signals, enhancing LLM spatial reasoning.  Experiments demonstrate that our method outperforms other methods in physical consistency and intent alignment, achieving automated construction from natural language to executable game scenes and providing a standardized solution for the democratization of AI Towns.

\section*{Limitations}
Although World Craft successfully achieves automated creation from natural language to executable game scenes, the current system still has limitations, which outline directions for future research.

\paragraph{1. Limitations on Scene Scale and Complexity.}
Current generation primarily focuses on indoor environments within single scenes (\eg, residences, offices, or interiors of single buildings). While the system can handle room layouts and component placement, it does not yet fully support complete ``town-level'' macroscopic planning that covers outdoor terrain, road networks, and multi-building coordination. Large-scale outdoor scenes involve more complex hierarchical structures, which is a key challenge we need to address in the next stage.

 \paragraph{2. Depth of Interaction Logic.}
While World Scaffold ensures basic physical interactability, the generated environments primarily support navigation, life simulation, and social activities. Current capabilities remain limited regarding advanced interaction logic involving complex physical simulations (\eg, fluids, destruction effects) or dynamic environmental evolution (\eg, constructing new environments in real-time during simulation).

\paragraph{Summary.}
In the future, we aim to extend World Craft to coordinated multi-scene open-world construction and further enrich asset diversity and interaction depth, thereby achieving truly fully automated AI Town generation.

\section*{Ethics Statement}
\paragraph{Data Privacy and Usage.} 
All training data utilized in this paper were constructed based on our proposed method. Detailed descriptions of the construction algorithms and referenced content are provided in the Appendix. Additionally, the APIs for both open-source and closed-source models employed in data generation are listed in the Appendix. Finally, the tile sets used in our asset library were sourced from open-source works by authors on professional asset platforms; detailed credits and URLs will be provided in our open-source release. All data have been anonymized to eliminate any personal or confidential information.

\paragraph{Human Evaluation Statement.} 
This study involves human subjects, and we strictly adhere to ethical guidelines to safeguard participant rights. Key measures include: (1) \textbf{Informed Consent}: Prior to the experiment, we fully disclosed the research objectives, procedures, and participant rights. Participants were informed of their freedom to withdraw from the study unconditionally at any stage without facing any adverse consequences. (2) \textbf{Data De-identification}: All evaluation data (including interaction logs and questionnaires) have undergone strict anonymization. By removing personal identifiers, we ensure that data cannot be traced back to specific individuals.

\bibliography{main}

\clearpage

\section*{Appendix}
\raggedbottom
\appendix
\label{sec:appendix}

\begin{figure*}[t]
    \centering
    \includegraphics[width=\linewidth]{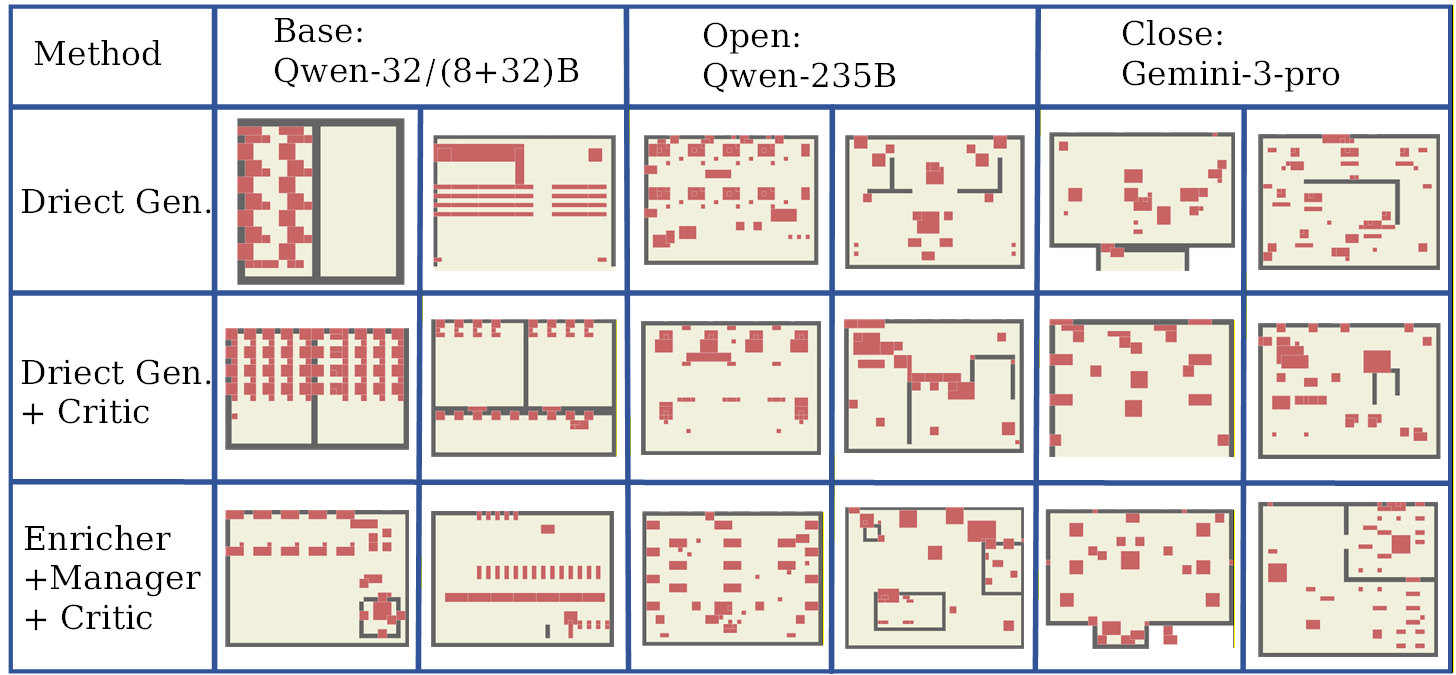}
    \caption{
        Examples of output results of the three models in three reasoning stages, Refer to the upper part of Table~\ref{tab:main_results}.        }
    \vspace{-4mm}
    \label{fig:baseline_evolution}
\end{figure*}

\begin{figure*}[t]
    \centering
    \includegraphics[width=0.9\linewidth]{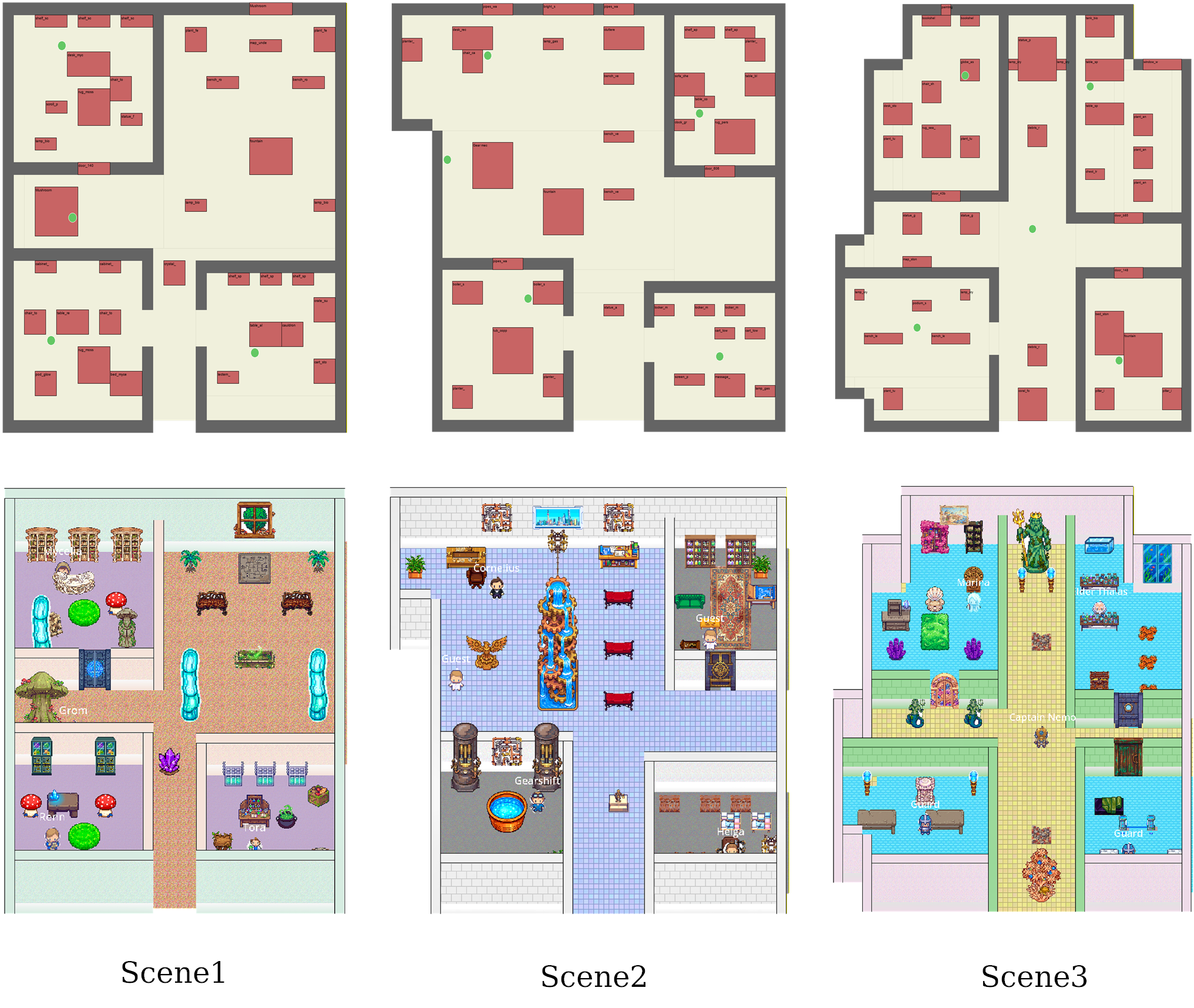}
    \caption{Examples of layout designs and final output results of our method in three scenarios.
        }
    \vspace{-4mm}
    \label{fig:ours_showcase}
\end{figure*}

\begin{figure*}[t]
    \centering
    \includegraphics[width=0.8\linewidth]{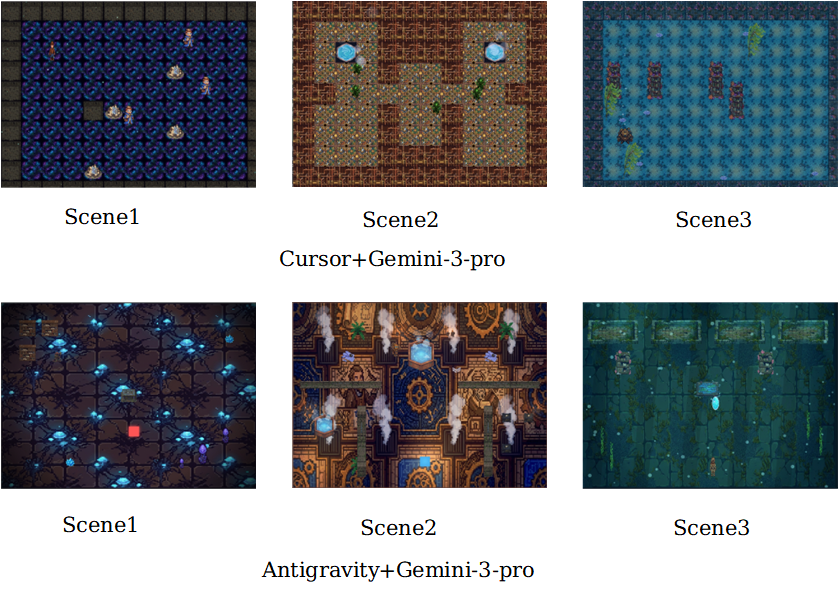}
    \caption{
        Examples of the performance of Cursor and Antigravity in three scenarios.
        }
    \vspace{-4mm}
    \label{fig:code_agents}
\end{figure*}

\section{Comparison of Model Output Results}
\label{sec:appA}
To intuitively demonstrate the performance differences between methods, we visualized the generated layouts for three diverse test scenarios: 

\textit{Scene 1 (A cool, glowing mycelium chamber holds ancient scrolls with floating spores and faint rustles.)};

\textit{Scene 2 (A luxurious underground bathhouse, where steam rises from hexagonal copper pools filled with mineral-rich water that shimmers with a faint blue glow. Exquisite mosaic tiles depict scenes of ancient inventors. Hidden panels on the walls slide open silently, revealing niches containing ticking lockboxes and half-burned blueprints.)};

\textit{Scene 3 (The once magnificent study is now submerged on the seabed. Its arched windows have shattered, obscured by swaying kelp. Coral has spread over the bookshelves and lectern, encasing the leather-bound classics in calcified, lace-like formations. Luminescent fish dart through floating clouds of ink, remnants of broken inkwells. On a large stone table lies a glowing slab inscribed with indecipherable symbols. Sunlight filters down from far above in fractured beams, illuminating drifting particles like underwater snowflakes.)}.

Fig~\ref{fig:baseline_evolution} first illustrates the generation results of three baseline models on Scene 2 and Scene 3 across the three stages defined in the upper part of Table~\ref{tab:main_results}. Observations indicate that in the absence of domain knowledge, relying solely on the multi-agent framework remains insufficient to effectively handle complex geometric constraints. In contrast, Fig.~\ref{fig:ours_showcase} presents the performance of our method (Ours in Table~\ref{tab:main_results}) across all scenarios, demonstrating the effectiveness of our data and domain knowledge injection. Finally, Fig.~\ref{fig:code_agents} compares the results of code agents (Cursor and Antigravity). These methods tend to construct environments with simplistic layouts and sparse elements, lacking visual expressiveness, thus failing to meet the requirements of ideal simulation environments.

\section{Experimental Parameters and Settings}
\label{sec:appB}
All training experiments were conducted on NVIDIA 8*H200 GPUs(141G). The detailed hyperparameter settings are listed in Table~\ref{tab:hyperparameters}.

\begin{table*}[!t]
\centering
\begin{tabular}{lcc}
\toprule
\textbf{Hyperparameter} & \textbf{Stage 1 (Enricher)} & \textbf{Stage 2 (Manager)} \\
\midrule
\multicolumn{3}{l}{\textit{Model Configuration}} \\
Base Model & Qwen3-8B & Qwen3-32B \\
Fine-tuning Method & Full-parameter & Full-parameter \\
Precision & bfloat16 & bfloat16 \\
Max Sequence Length & 2048 & 12000 \\
DeepSpeed Strategy & ZeRO-3 & ZeRO-3 \\
\midrule
\multicolumn{3}{l}{\textit{Training Optimization}} \\
Optimizer & AdamW & AdamW \\
Learning Rate (lr) & 2.0e-5 & 1.0e-5 \\
LR Scheduler Type & Cosine & Cosine \\
Warmup Ratio & 0.1 & 0.1 \\
Num. Epochs & 5.0 & 5.0 \\
Batch Size (per device) & 4 & 1 \\
Gradient Accum. Steps & 4 & 1 \\
Total Batch Size\footnotemark[1] & 128 & 8 \\
\bottomrule
\end{tabular}
\caption{Detailed Hyperparameters for Two-Stage Instruction Tuning.}
\label{tab:hyperparameters}
\end{table*}

\clearpage

\section{Details of Asset Library}
\label{sec:appC}

\begin{figure*}[t]
    \centering
    \includegraphics[width=\linewidth]{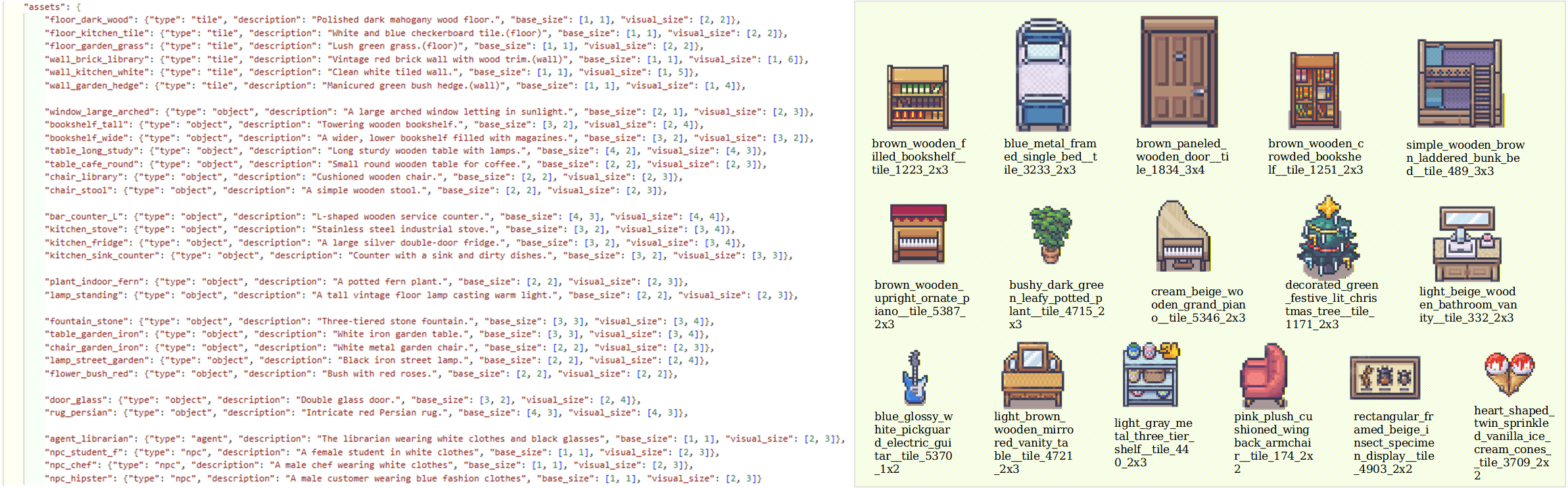}
    \caption{
    \textbf{Examples of entries in the Asset Library ($\mathcal{D}_{lib}$).} We utilize open-source tile sets as the foundation for our localized library. Due to copyright restrictions, we present only a subset of representative examples here. Full credits and links to the original artists will be provided after open-sourcing.
        }
    \vspace{-4mm}
    \label{fig:assest}
\end{figure*}

To address style fragmentation in text-to-image generation, we constructed a asset library(5500+), as shown in Fig.~\ref{fig:assest}. The retrieval process aims to identify the most relevant reference image  that matches both the semantic content and spatial dimensions of the target component. This algorithm employs a two-stage strategy: Token Matching and Dimension Ranking. As outlined in Algorithm~\ref{alg:asset_retrieval}, the system first filters candidate assets based on keyword intersection. Subsequently, it calculates the Manhattan distance between the target dimensions and candidate tile dimensions to minimize spatial distortion.

\begin{algorithm}[h]
    \caption{Asset Retrieval Strategy}
    \label{alg:asset_retrieval}
    \SetKwInOut{Input}{Input}
    \SetKwInOut{Output}{Output}
    \SetKwFunction{Normalize}{Normalize}
    \SetKwFunction{Intersection}{Intersection}
    \Input{Target Asset ID $S_{id}$, Description $S_{desc}$, Target Dims $D_{target}(w,h)$, Index $\mathcal{D}_{lib}$}
    \Output{Best Matching Reference Image $v_{ref}$}
    \tcp{Step 1: Query Normalization}
    $Q_{text} \leftarrow S_{id} \oplus \text{`` ''} \oplus S_{desc}$\;
    $Q_{tokens} \leftarrow$ \Normalize{$Q_{text}$}\tcp*{Remove stop words}
    $v_{ref} \leftarrow \text{None}$\;
    $P_{min} \leftarrow \infty$\tcp*{Initialize penalty}
    \tcp{Step 2: Traverse Asset Library}
    \For{each candidate asset $A_i \in \mathcal{D}_{lib}$}{
        $T_{i} \leftarrow A_i.\text{tokens}$\;
        \tcp{Stage 1: Semantic Filtering}
        \If{\Intersection{$Q_{tokens}, T_{i}$} $\neq \emptyset$}{
            $D_{i} \leftarrow A_i.\text{dimensions}$\;
            \tcp{Stage 2: Dimension Ranking}
            $P_{curr} \leftarrow |D_{target}.w - D_{i}.w| + |D_{target}.h - D_{i}.h|$\;
            \If{$P_{curr} < P_{min}$}{
                $P_{min} \leftarrow P_{curr}$\;
                $v_{ref} \leftarrow A_i.\text{path}$\;
                \If{$P_{min} == 0$}{
                    \textbf{break}\tcp*{Perfect match}
                }
            }
        }
    }
    \KwRet{$v_{ref}$}
\end{algorithm}

\clearpage

\section{Procedural Generation Logic and Data-Driven Priors}
\label{sec:appH}
To ensure that the synthetic layouts generated in Stage 1 structurally align with real-world patterns, we extracted a set of architectural priors from the RPLAN dataset~\cite{56_10.1145/3355089.3356556}. Given that RPLAN is strictly limited to residential floor plans, whereas our framework aims to construct general-purpose agent environments covering diverse functions (\eg, offices, retail), we focused on extracting generalizeable geometric and topological rules rather than relying directly on domain-specific samples. This strategy enables us to generate infinite and diverse scenes using limited data. Below are examples of key constraints applied in the pipeline:
\paragraph{Geometric Orthogonality.} Observations indicate that the vast majority of wall segments are axis-aligned. Therefore, our generation algorithm enforces a strict orthogonality constraint and operates on a discrete grid. This ensures all walls remain horizontal or vertical, preventing irregular angles and ensuring the generated structures comply with general architectural norms.
\paragraph{Topological Centrality.} Real-world layouts typically evolve around public spaces. To simulate this topological feature, the algorithm adopts a ``periphery-to-center partitioning'' logic: it first initializes the overall building envelope and then iteratively carves private rooms from the boundary inwards. The remaining unpartitioned area naturally forms the central public core, mathematically guaranteeing a connected backbone structure and avoiding isolated regions.
\paragraph{Boundary Morphology.} To simulate complex contours formed in real buildings due to lighting or zoning requirements, our algorithm implements a shape grammar. This iteratively augments the initial building envelope with randomized sub-structures, producing complex non-convex boundaries that more closely resemble real-world floor plan footprints.
\paragraph{Dimensional Calibration.} We constrain generation using aspect ratio and area thresholds derived from real-world distributions, rather than using random parameters. This ensures that every generated space is physically usable, preventing the creation of geometrically valid but functionally uninhabitable ``splinter'' corners.
\paragraph{Path Optimization.} Door placement determines circulation efficiency. When connecting rooms, our algorithm employs a distance-minimization cost function to automatically locate wall segments that minimize the distance to the central area. This placement strategy effectively replicates the efficient circulation patterns found in human-designed floor plans.
\paragraph{Wall Continuity and Thickness.} To align with RPLAN annotation standards (where walls are represented with consistent pixel thickness), our generator includes a geometric regularization step. This step merges fragmented boundary segments into continuous geometric primitives, ensuring the result is topologically equivalent to the semantic segmentation maps provided in the original dataset.

The pseudocode for the synthetic layout generation step is shown in Algorithm~\ref{alg:layout_synthesis}.

\begin{algorithm}[!ht]
    \caption{RPLAN-Aligned Layout Synthesis}
    \label{alg:layout_synthesis}
    \SetKwInOut{Input}{Input}
    \SetKwInOut{Output}{Output}
    \SetKwFunction{InitSeed}{InitSeed}
    \SetKwFunction{Augment}{AugmentShape}
    \SetKwFunction{Scan}{ScanRegion}
    \SetKwFunction{Carve}{PartitionRoom}
    \SetKwFunction{Dist}{Dist}
    \SetKwFunction{Centroid}{Centroid}
    \SetKwFunction{Regularize}{RegularizeWalls}

    \Input{Target Room Count $N_{rooms}$, Map Dims $W, H$, RPLAN Priors $\mathbf{\Theta}$}
    \Output{Semantic Grid Layout $G$}

    \tcp{Step 1: Non-Convex Boundary Generation}
    $S_{core} \leftarrow$ \InitSeed{$W, H$}\;
    $B_{poly} \leftarrow$ \Augment{$S_{core}, \mathbf{\Theta}_{shape}$}\tcp*{Add sub-rects per Shape Grammar}
    $G \leftarrow \text{Rasterize}(B_{poly})$\;

    \tcp{Step 2: Topology-Preserving Partitioning}
    $Q_{corners} \leftarrow \text{GetVertices}(B_{poly})$\;
    $R_{list} \leftarrow \emptyset$\;
    \While{$|R_{list}| < N_{rooms} \textbf{ and } Q_{corners} \neq \emptyset$}{
        $c \leftarrow \text{PopRandom}(Q_{corners})$\;
        \tcp{Constraint: Dimensional Calibration}
        $R_{cand} \leftarrow$ \Scan{$c, \mathbf{\Theta}_{dim}$}\;
        \If{$\text{IsValid}(R_{cand}) \textbf{ and } \text{IsConnected}(G \setminus R_{cand})$}{
            $G \leftarrow$ \Carve{$G, R_{cand}$}\;
            $R_{list}.\text{append}(R_{cand})$\;
        }
    }

    \tcp{Step 3: Circulation Optimization}
    $C_{core} \leftarrow$ \Centroid{$S_{core}$}\;
    \For{each room $r \in R_{list}$}{
        $W_{cand} \leftarrow \text{FindValidWallSegments}(r)$\;
        \tcp{Minimize distance to functional core}
        $p_{best} \leftarrow \arg\min_{p \in W_{cand}} \Dist(p, C_{core})$\;
        $G[p_{best}] \leftarrow \text{DOOR}$\;
    }

    \tcp{Step 4: Geometric Regularization}
    $G \leftarrow$ \Regularize{$G$}\tcp*{Merge segments \& uniform thickness}
    
    \KwRet{$G$}
\end{algorithm}

\clearpage

\section{Details of manual evaluation}
\label{sec:appD}
\paragraph{Evaluator Profile} To ensure professional judgment regarding game scene layouts, we recruited five independent evaluators for the reviews in Sections~\ref{sec:codeganet} and~\ref{sec:humaneval}. All participants hold at least a Bachelor's degree, possess an average of over five years of experience in RPG or simulation strategy games, and are familiar with common game map mechanics and navigation logic. Additionally, for the Code Agent (Cursor, Antigravity) operation tasks in Section~\ref{sec:codeganet}, the three operators were doctoral students specializing in Artificial Intelligence, ensuring the standardized use of the tools.

\paragraph{Questionnaire Design} To minimize cognitive load during the evaluation, we synthesized the eight automated metrics from the main text into three dimensions of pairwise forced-choice questions. Evaluators were presented with the input instruction and two generated scene images, and asked to make decisions based on the following criteria:

\textbf{Layout Plausibility} (Correlating with CFR, RCS, OPS): ``Which scene's layout is physically more reasonable and visually more harmonious?'' This dimension assesses spatial connectivity and the logical placement of objects. 

\textbf{Content Richness} (Correlating with CER, OVD, PAC): ``Which scene contains more valid details and fewer incongruous objects?'' This dimension focuses on asset diversity and visual artifacts such as floating or overlapping objects. 

\textbf{Intent Consistency} (Correlating with VSA): ``Which scene more accurately reflects the input text description?'' This dimension evaluates the semantic fidelity of the generated result to the natural language instruction.

\section{Details of Evaluation Metrics}
\label{sec:appE}
This section compiles the specific prompts utilized for the VLM-based scoring metrics discussed in the main text. To ensure reproducibility and transparency, we present the full content of the instructions input to the \textbf{Gemini-3-Pro} model. The detailed prompts for Object Placement Reasonableness (OPS), Object Volume Density (OVD), Physical Attribute Consistency (PAC), and Visual Harmony (VH) are illustrated in Fig.~\ref{fig:prompt_ops}, Fig.~\ref{fig:prompt_ovd}, Fig.~\ref{fig:prompt_pac}, and Fig.~\ref{fig:prompt_vh}, respectively.

\clearpage
\onecolumn

\begin{tcolorbox}[
    enhanced,
    breakable,            
    colback=white,           
    colframe=gray!60!black,   
    coltitle=white,          
    title=\textbf{Prompt for Object Placement Reasonableness (OPS)},
    arc=0pt, outer arc=0pt,
    boxrule=1pt,
    top=5pt, bottom=5pt,
]
    \begin{lstlisting}[breaklines=true, basicstyle=\small\ttfamily, columns=fullflexible, extendedchars=false]
        This is a design document for a game scene. Please read it carefully, and your task is to identify semantically unreasonable elements in the scene.
        Scene Description: "{scene_desc}"
        Complete Original Design Data of the Scene:{design_data_str}
        Data Reading Guidelines
            1.Position Information: Check the object_layer (including position coordinates) under the layout field, as well as the wall_layer/floor_layer (including area regions).
            2.Object Attributes: Use the asset_id to find the corresponding object name (name) and size (base_size) in the assets field.
            3.Task: Judge the design logic by combining coordinate positions and object attributes.
        
        Please identify objects that are logically incompatible or extremely unusual in the context of this specific scene.(Examples: A toilet should never appear in a kitchen or main seating area such as a living room corridor; A main entrance must not be blocked by large obstacles like bookshelves or pianos, unless it is an intentional decorative screen.)
        Coordinate-Based Judgment Required: Check if objects are located in incorrect room areas (e.g., A toilet falls within the kitchen floor area based on coordinate calculations).

        You must return a JSON object:
        {{
            "unreasonable_objects": ["Object name1", "Object name2"],
            "count": 2,
            "reason": "Brief explanation (if position errors are involved, please reference coordinates for explanation)"
        }}
        If all elements are reasonable, set "count": 0.
        Do not output any extra content, especially greetings or irrelevant remarks.
    \end{lstlisting}
\end{tcolorbox}

\begin{center}
    \captionsetup{hypcap=false}
    \captionof{figure}{Evaluation Prompt for Object Placement Reasonableness (OPS).}
    \label{fig:prompt_ops}
\end{center}

\begin{tcolorbox}[
    enhanced,
    breakable,            
    colback=white,           
    colframe=gray!60!black,   
    coltitle=white,          
    title=\textbf{Prompt for Object Volume Density (OVD)},
    arc=0pt, outer arc=0pt,
    boxrule=1pt,
    top=5pt, bottom=5pt,
]
    \begin{lstlisting}[breaklines=true, basicstyle=\small\ttfamily, columns=fullflexible, extendedchars=false]
        Role: You are an interior design critic.
        Task: Evaluate the "object density" and "spatial suitability" of the provided floor plan image.Scene Background: "{scene_desc}"
        The image shows a layout: Green = floor, Red = objects/obstacles, Black = empty space.
        Detailed Scoring Rubric (0-10):
            [0-2] Severe Failure (Unusable)
            Severe clutter: Objects completely block the main entrance or passageways. No navigable paths exist.
            OR Extreme emptiness: Despite being described as a functional space, the room is essentially vacant.
            [3-4] Poor (Imbalanced/Awkward)
            Poor circulation: Paths exist but are zigzag, overly narrow, or cramped.
            Uneven distribution: All objects cluster in one corner, leaving the rest as unused space.
            Proportion issues: Objects appear excessively large or small relative to the room.
            [5-6] Fair (Functional but Mediocre)
            Navigable: Passage is possible, and furniture is placed.
            Lack of design sense: Looks randomly arranged or a basic grid layout, with no consideration for aesthetics or functional zones.
            [7-8] Good (Professional Layout)
            Clear zoning: Distinct areas designated for specific activities.
            Excellent circulation: Retains clear, spacious traffic paths (main thoroughfares).
            Balanced density: Objects are reasonably distributed in the space, without causing a sense of overcrowding.
            [9-10] Excellent (Masterful/Perfect)
            Perfect context alignment: Density perfectly matches the scene description (e.g., a "cozily cluttered study" feels cozy rather than messy; a "minimalist gallery" feels open rather than empty).
            Negative space: Exceptional use of empty space ("breathing room") to complement objects.
            Natural composition: The layout feels natural and lived-in, not computer-generated.
        Output JSON ONLY:
        {{
            "score": 8,
            "reason": "The cafe has distinct zones and good flow."
        }}
    \end{lstlisting}
\end{tcolorbox}

\begin{center}
    \captionsetup{hypcap=false}
    \captionof{figure}{Evaluation Prompt for Object Volume Density (OVD).}
    \label{fig:prompt_ovd}
\end{center}

\begin{tcolorbox}[
    enhanced,
    breakable,            
    colback=white,           
    colframe=gray!60!black,   
    coltitle=white,          
    title=\textbf{Prompt for Physical Attribute Consistency (PAC)}, 
    arc=0pt, outer arc=0pt,
    boxrule=1pt,
    top=5pt, bottom=5pt,
]
    \begin{lstlisting}[breaklines=true, basicstyle=\small\ttfamily, columns=fullflexible, extendedchars=false]
        This is the size of all elements in a game scene. You need to check the physical attributes and size consistency of these elements to see if their sizes are reasonable.
        Below is a list of assets used in the scene, including:
        Base size: The grid footprint on the floor (e.g., [1, 1] represents a 1x1 grid). For example, a refrigerator has a footprint of 2x1, and a table has a footprint of 2x2.
        Visual size: The actual visual size of the object. For objects with low or no height (such as carpets), their visual size must be equal to the base size. For objects with a certain height or considerable height, the height of the visual size is often one, two, or three units higher than the base size.
        For example, a table with a base size of 2x2 may have a visual size of 2x3; a refrigerator with a base size of 2x1 may have a visual size of 2x4.
        Here is the data content:{data_str}
        Evaluation Criteria:
            1.Footprint logic: Is the base size reasonable? (e.g., a "grand piano" cannot be 1x1).
            2.Vertical proportion and size (critical): Check the visual_size (especially the height).
              -Relative proportion: Compare objects with each other.
              -Violation example: If the visual height of a "chair" is 5.0 while the height of a "table" is 1.0, this is a serious violation.
              -Violation example: A "refrigerator" should not be smaller than a "stool".

        You need to return a JSON object:
        {{
            "violation_objects": ["name_of_bad_object_1", "name_of_bad_object_2"],
            "count": <int>,
            "reason": "Specific explanation referencing the dimensions (e.g., 'Chair height 5.0 is disproportionate to Table height 1.2')."
        }}
        If everything is reasonable, return "count": 0.
        Do not output any other extra content, especially various greeting contents.
    \end{lstlisting}
\end{tcolorbox}

\begin{center}
    \captionsetup{hypcap=false}
    \captionof{figure}{Evaluation Prompt for Physical Attribute Consistency (PAC).}
    \label{fig:prompt_pac}
\end{center}

\vspace{-3mm}
\begin{tcolorbox}[
    enhanced,
    breakable,            
    colback=white,            
    colframe=gray!60!black,   
    coltitle=white,           
    title=\textbf{Prompt for Pairwise Preference Evaluation}, 
    arc=0pt, outer arc=0pt,
    boxrule=1pt,
    top=5pt, bottom=5pt,
]
    \begin{lstlisting}[breaklines=true, basicstyle=\small\ttfamily, columns=fullflexible, extendedchars=false]
    Here are two AI-generated town environments based on the textual description: "{text}".
    Please evaluate them considering the following criteria:
    1. Instruction Alignment: Does the scene accurately include the elements and atmosphere described in the text?
    2. Visual Aesthetics: Is the artistic style consistent, detailed, and visually appealing?
    3. Layout & Playability: Is the spatial structure logical and inviting for a player to explore?
    Based on a comprehensive assessment of these factors, you only need to answer the final question: In which game environment would you prefer to play? The output result is 1 or 2.

    \end{lstlisting}
\end{tcolorbox}

\begin{center}
    \captionsetup{hypcap=false}
    \captionof{figure}{Evaluation Prompt for Pairwise Preference (Human \& VLM).}
    \label{fig:prompt_preference} 
\end{center}

\vspace{-3mm}
\begin{tcolorbox}[
    enhanced,
    breakable,            
    colback=white,           
    colframe=gray!60!black,   
    coltitle=white,          
    title=\textbf{Prompt for Visual Harmony (VH)}, 
    arc=0pt, outer arc=0pt,
    boxrule=1pt,
    top=5pt, bottom=5pt,
]
    \begin{lstlisting}[breaklines=true, basicstyle=\small\ttfamily, columns=fullflexible, extendedchars=false]
    Role: You are an Art Director for a top-tier game studio.
    Task: Evaluate the "Stylistic Consistency" and "Visual Harmony" of the generated game scene image.
    Context: This scene is constructed from various tiles and resources. Your goal is to strictly determine whether these elements blend into a unified artistic whole or appear disjointed.
    Evaluation Dimensions:
    1. Artistic Consistency: Do all assets share the same rendering style (e.g., pixel art resolution, stroke style, level of detail)?
    2. Color Harmony: Is there a unified color palette and lighting temperature across the scene?
    3. Integration: Do objects look naturally grounded, or do they look like stickers pasted onto a background (artifacts, jagged edges, mismatched perspective)?

    Detailed Scoring Rubric (0-10):
    [0-2] Visual Chaos (Frankenstein-like)
    Severe mismatch: The scene looks like a random collage. For example, a photorealistic chair placed next to a 8-bit pixel art table.
    Clashing palettes: Colors are jarring and uncoordinated.
    
    [3-4] Disjointed (Lack of Polish)
    Noticeable inconsistency: Most items match, but 3-4 objects clearly belong to a different art style or asset pack.
    Poor blending: Visible seams between tiles or distinct lighting differences between objects and the floor.
    
    [5-6] Acceptable (Generic Consistency)
    Unified but dull: The style is generally consistent, but looks flat or mechanical.
    Minor flaws: Slight variations in pixel density or color saturation, but nothing scene-breaking.
    
    [7-8] Good (Cohesive Style)
    Strong theme: All elements clearly belong to the same specific world (e.g., "Cyberpunk" or "Victorian").
    Harmonious colors: The color palette is pleasing and consistent. Objects interact well with the background.
    
    [9-10] Masterpiece (Perfect Unity)
    Seamless integration: It is impossible to tell individual assets apart; the scene looks like a single, hand-painted illustration.
    Atmospheric unity: Lighting, shadows, and textures work together to create a compelling mood.

You must return a JSON object:
{{
    "score": <int>,
    "reason": "Specific critique (e.g., 'The vending machine has a different pixel resolution than the wall, breaking immersion.')"
}}
Do not output any extra content.
    \end{lstlisting}
\end{tcolorbox}

\begin{center}
    \captionsetup{hypcap=false}
    \captionof{figure}{Evaluation Prompt for Visual Harmony (VH).}
    \label{fig:prompt_vh}
\end{center}

\clearpage
\twocolumn

\section{Dataset Construction and Statistics}
\label{sec:appF}
\begin{figure*}[t]
    \centering
    \includegraphics[width=\linewidth]{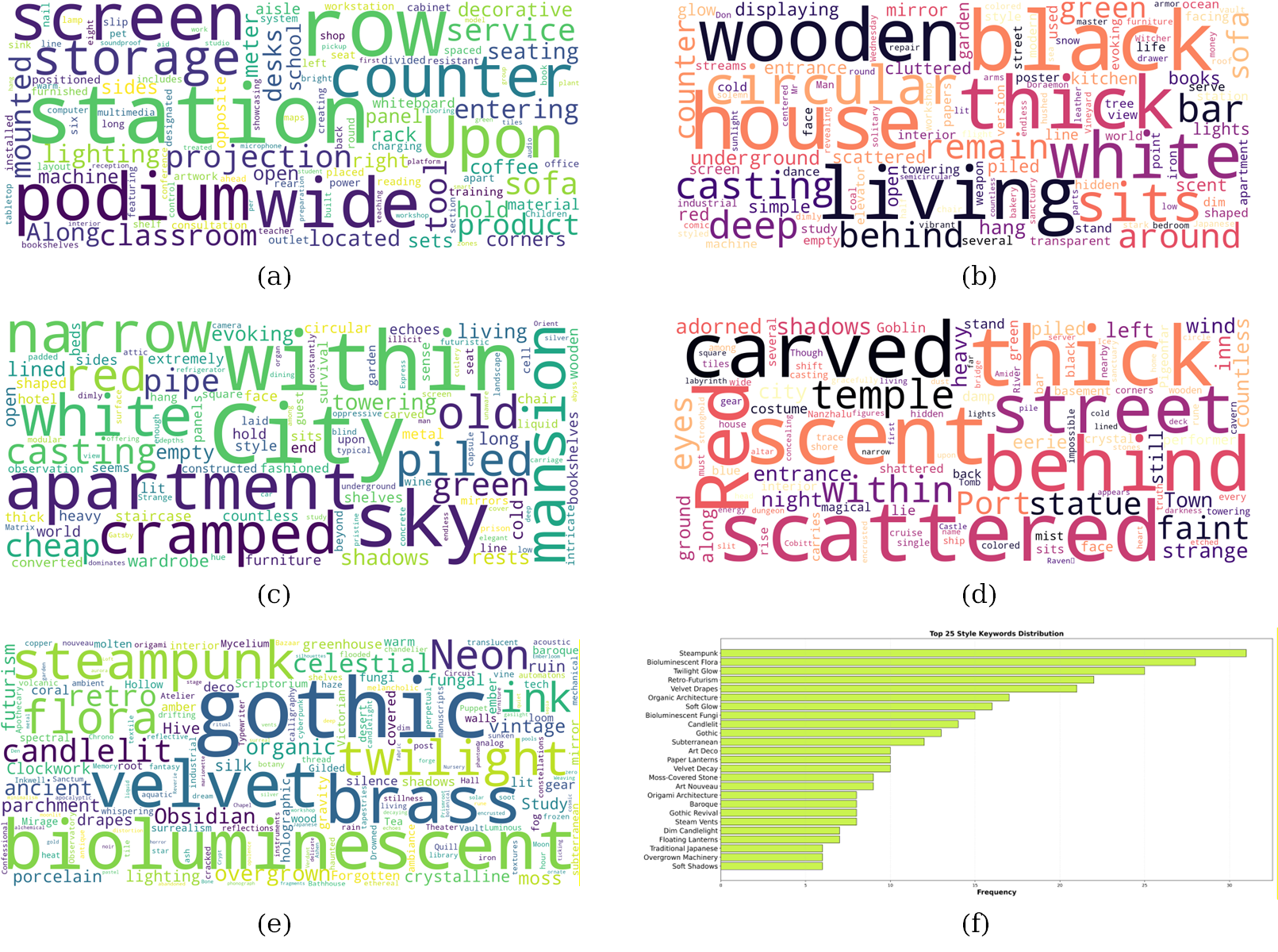}
    \caption{
        Dataset statistics and distribution. 
        (a)-(d) Semantic word clouds for the four initial data collection domains. 
        (e) Vocabulary distribution of the final augmented dataset. 
        (f) The frequency distribution of the Top-25 style categories.
    }
    \vspace{-4mm}
    \label{fig:dataset_stat}
\end{figure*}

To complement the general description in Section~\ref{sec:data}, this section details the construction sources and distribution characteristics of the dataset. Fig.s~\ref{fig:dataset_stat}(a)-(d) first visualize the semantic word clouds from the four collection sources of the raw data-Real-world Scenarios, Literature, Film \& TV, and TRPG Games-highlighting the diversity of the initial corpus. Subsequently, Fig.~\ref{fig:dataset_stat}(e) presents the vocabulary distribution of the final dataset (2,000 samples) after stylistic augmentation based on the training set (400 samples). Finally, Fig.~\ref{fig:dataset_stat}(f) illustrates the distribution of the Top-25 style categories in the augmented dataset.

\section{Data Structure Definitions and Examples}
\label{sec:appG}
To complement the content in Section 3.1, this section provides a complete data example of the structured layout $\mathcal{G}$, as shown in Fig.~\ref{fig:layout}. This example specifically illustrates the internal details of the quadruple $(M, A, L, P)$, including Metadata ($M$) defining basic scene configurations, Asset Definitions ($A$) describing visual styles and layer attributes, Layout ($L$) establishing precise spatial topology, and Properties ($P$) specifying collision and interaction logic, intuitively demonstrating how the model instantiates the generation target into executable game environment data.

\clearpage
\onecolumn

\begin{tcolorbox}[
    enhanced,
    breakable,            
    colback=white,           
    colframe=gray!60!black,   
    coltitle=white,          
    title=\textbf{Structured Layout Output ($\mathcal{G}$)},
    arc=0pt, outer arc=0pt,
    boxrule=1pt,
    top=5pt, bottom=5pt,
]
    \begin{lstlisting}[breaklines=true, basicstyle=\small\ttfamily, columns=fullflexible, extendedchars=false]
    "metadata": {
        "scene_name": "The Verdant Scholar's Hub",
        "grid_size": [35, 28],
        "description": "A bustling establishment. The interior is subdivided into a quiet reading zone and a social cafe area. The kitchen is fully equipped. The garden is romantic and well-lit.",
        "style_prompt": "16-bit pixel art, top-down RPG style, cozy lighting, clutter and details"
    },
    "assets": {
        "floor_dark_wood": {"type": "tile", "description": "Polished dark mahogany wood floor.", "base_size": [1, 1], "visual_size": [2, 2]},
        "floor_kitchen_tile": {"type": "tile", "description": "White and blue checkerboard tile.(floor)", "base_size": [1, 1], "visual_size": [2, 2]},
        "floor_garden_grass": {"type": "tile", "description": "Lush green grass.(floor)", "base_size": [1, 1], "visual_size": [2, 2]},
        "wall_brick_library": {"type": "tile", "description": "Vintage red brick wall with wood trim.(wall)", "base_size": [1, 1], "visual_size": [1, 6]},
        "wall_kitchen_white": {"type": "tile", "description": "Clean white tiled wall.", "base_size": [1, 1], "visual_size": [1, 5]},
        "wall_garden_hedge": {"type": "tile", "description": "Manicured green bush hedge.(wall)", "base_size": [1, 1], "visual_size": [1, 4]},
        "window_large_arched": {"type": "object", "description": "A large arched window letting in sunlight.", "base_size": [2, 1], "visual_size": [2, 3]},
        "bookshelf_tall": {"type": "object", "description": "Towering wooden bookshelf.", "base_size": [3, 2], "visual_size": [2, 4]},
        "bookshelf_wide": {"type": "object", "description": "A wider, lower bookshelf filled with magazines.", "base_size": [3, 2], "visual_size": [3, 2]},
        "table_long_study": {"type": "object", "description": "Long sturdy wooden table with lamps.", "base_size": [4, 2], "visual_size": [4, 3]},
        "table_cafe_round": {"type": "object", "description": "Small round wooden table for coffee.", "base_size": [2, 2], "visual_size": [2, 3]},
        "chair_library": {"type": "object", "description": "Cushioned wooden chair.", "base_size": [2, 2], "visual_size": [2, 3]},
        "chair_stool": {"type": "object", "description": "A simple wooden stool.", "base_size": [2, 2], "visual_size": [2, 3]},
        "bar_counter_L": {"type": "object", "description": "L-shaped wooden service counter.", "base_size": [4, 3], "visual_size": [4, 4]},
        "kitchen_stove": {"type": "object", "description": "Stainless steel industrial stove.", "base_size": [3, 2], "visual_size": [3, 4]},
        "kitchen_fridge": {"type": "object", "description": "A large silver double-door fridge.", "base_size": [3, 2], "visual_size": [3, 4]},
        "kitchen_sink_counter": {"type": "object", "description": "Counter with a sink and dirty dishes.", "base_size": [3, 2], "visual_size": [3, 3]},
        "plant_indoor_fern": {"type": "object", "description": "A potted fern plant.", "base_size": [2, 2], "visual_size": [2, 3]},
        "lamp_standing": {"type": "object", "description": "A tall vintage floor lamp casting warm light.", "base_size": [2, 2], "visual_size": [2, 3]},
        "fountain_stone": {"type": "object", "description": "Three-tiered stone fountain.", "base_size": [3, 3], "visual_size": [3, 4]},
        "table_garden_iron": {"type": "object", "description": "White iron garden table.", "base_size": [3, 3], "visual_size": [3, 4]},
        "chair_garden_iron": {"type": "object", "description": "White metal garden chair.", "base_size": [2, 2], "visual_size": [2, 3]},
        "lamp_street_garden": {"type": "object", "description": "Black iron street lamp.", "base_size": [2, 2], "visual_size": [2, 4]},
        "flower_bush_red": {"type": "object", "description": "Bush with red roses.", "base_size": [2, 2], "visual_size": [2, 2]},
        "door_glass": {"type": "object", "description": "Double glass door.", "base_size": [3, 2], "visual_size": [2, 4]},
        "rug_persian": {"type": "object", "description": "Intricate red Persian rug.", "base_size": [4, 3], "visual_size": [4, 3]},
        "agent_librarian": {"type": "agent", "description": "The librarian wearing white clothes and black glasses", "base_size": [1, 1], "visual_size": [2, 3]},
        "npc_student_f": {"type": "npc", "description": "A female student in white clothes", "base_size": [1, 1], "visual_size": [2, 3]},
        "npc_chef": {"type": "npc", "description": "A male chef wearing white clothes", "base_size": [1, 1], "visual_size": [2, 3]},
        "npc_hipster": {"type": "npc", "description": "A male customer wearing blue fashion clothes", "base_size": [1, 1], "visual_size": [2, 3]}
    },
    "layout": {
        "floor_layer": [
            {"asset_id": "floor_dark_wood", "command": "fill_rect", "area": [0, 0, 25, 20]},
            {"asset_id": "floor_kitchen_tile", "command": "fill_rect", "area": [25, 0, 10, 20]},
            {"asset_id": "floor_garden_grass", "command": "fill_rect", "area": [0, 20, 35, 8]}
        ],
        "wall_layer": [
            {"asset_id": "wall_brick_library", "command": "fill_rect", "area": [0, 0, 35, 1]},
            {"asset_id": "wall_brick_library", "command": "fill_rect", "area": [0, 1, 1, 19]},
            {"asset_id": "wall_brick_library", "command": "fill_rect", "area": [34, 1, 1, 19]},
            {"asset_id": "wall_kitchen_white", "command": "fill_rect", "area": [24, 1, 1, 10]},
            {"asset_id": "wall_garden_hedge", "command": "fill_rect", "area": [0, 19, 11, 1]},
            {"asset_id": "wall_garden_hedge", "command": "fill_rect", "area": [14, 19, 21, 1]},
            {"asset_id": "wall_garden_hedge", "command": "fill_rect", "area": [0, 20, 1, 7]},
            {"asset_id": "wall_garden_hedge", "command": "fill_rect", "area": [34, 20, 1, 7]},
            {"asset_id": "wall_garden_hedge", "command": "fill_rect", "area": [0, 27, 35, 1]}
        ],
        "object_layer": [

            { "asset_id": "window_large_arched", "position": [3, 0] },
            { "asset_id": "window_large_arched", "position": [10, 0] },
            { "asset_id": "window_large_arched", "position": [17, 0] },

            { "asset_id": "bookshelf_tall", "position": [1, 1] },
            { "asset_id": "bookshelf_tall", "position": [5, 1] },
            { "asset_id": "bookshelf_tall", "position": [1, 6] },
            { "asset_id": "bookshelf_wide", "position": [5, 6] },

            { "asset_id": "rug_persian", "position": [6, 12] },
            { "asset_id": "table_long_study", "position": [6, 12] },
            { "asset_id": "chair_library", "position": [4, 12] },
            { "asset_id": "chair_library", "position": [5, 12] },
            { "asset_id": "chair_library", "position": [8, 12] },
            { "asset_id": "chair_library", "position": [9, 12] },
            { "asset_id": "lamp_standing", "position": [3, 11] },

            { "asset_id": "table_cafe_round", "position": [15, 10] },
            { "asset_id": "chair_library", "position": [14, 10] },
            { "asset_id": "chair_library", "position": [17, 10] },
            { "asset_id": "table_cafe_round", "position": [15, 15] },
            { "asset_id": "chair_library", "position": [14, 15] },
            { "asset_id": "chair_library", "position": [17, 15] },
            { "asset_id": "plant_indoor_fern", "position": [18, 8] },

            { "asset_id": "bar_counter_L", "position": [20, 4] },
            { "asset_id": "chair_stool", "position": [20, 7] },
            { "asset_id": "chair_stool", "position": [21, 7] },

            { "asset_id": "kitchen_stove", "position": [26, 1] },
            { "asset_id": "kitchen_fridge", "position": [30, 1] },
            { "asset_id": "kitchen_sink_counter", "position": [32, 4] },

            { "asset_id": "door_glass", "position": [12, 20] },

            { "asset_id": "fountain_stone", "position": [28, 23] },
            { "asset_id": "flower_bush_red", "position": [26, 22] },
            { "asset_id": "flower_bush_red", "position": [32, 22] },

            { "asset_id": "table_garden_iron", "position": [6, 23] },
            { "asset_id": "chair_garden_iron", "position": [5, 23] },
            { "asset_id": "chair_garden_iron", "position": [8, 23] },
            { "asset_id": "table_garden_iron", "position": [15, 23] },
            { "asset_id": "chair_garden_iron", "position": [14, 23] },
            { "asset_id": "chair_garden_iron", "position": [17, 23] },
            { "asset_id": "lamp_street_garden", "position": [10, 22] },
            { "asset_id": "lamp_street_garden", "position": [20, 22] }
        ],
        "npc_layer": [
            { "asset_id": "agent_librarian", "position": [21, 5] },
            { "asset_id": "npc_student_f", "position": [5, 13] },
            { "asset_id": "npc_chef", "position": [28, 3] },
            { "asset_id": "npc_hipster", "position": [14, 11] }
        ]
    },
    "properties": {
        "floor_dark_wood": { "physics": "passable", "navigation": "walkable", "semantic_tag": "floor_main" },
        "floor_kitchen_tile": { "physics": "passable", "navigation": "walkable", "semantic_tag": "floor_kitchen" },
        "floor_garden_grass": { "physics": "passable", "navigation": "walkable", "semantic_tag": "floor_garden" },
        "wall_brick_library": { "physics": "solid", "navigation": "obstacle", "semantic_tag": "wall" },
        "wall_kitchen_white": { "physics": "solid", "navigation": "obstacle", "semantic_tag": "wall" },
        "wall_garden_hedge": { "physics": "solid", "navigation": "obstacle", "semantic_tag": "fence" },
        "window_large_arched": { "physics": "solid", "navigation": "obstacle", "semantic_tag": "window" },
        "bookshelf_tall": { "physics": "solid", "navigation": "obstacle", "semantic_tag": "bookshelf" },
        "bookshelf_wide": { "physics": "solid", "navigation": "obstacle", "semantic_tag": "bookshelf" },
        "table_long_study": { "physics": "solid", "navigation": "obstacle", "semantic_tag": "table_study" },
        "table_cafe_round": { "physics": "solid", "navigation": "obstacle", "semantic_tag": "table_cafe" },
        "kitchen_stove": { "physics": "solid", "navigation": "obstacle", "semantic_tag": "stove" },
        "kitchen_fridge": { "physics": "solid", "navigation": "obstacle", "semantic_tag": "fridge" },
        "kitchen_sink_counter": { "physics": "solid", "navigation": "obstacle", "semantic_tag": "sink" },
        "bar_counter_L": { "physics": "solid", "navigation": "obstacle", "semantic_tag": "counter" },
        "chair_library": { "physics": "passable", "navigation": "obstacle", "semantic_tag": "chair" },
        "chair_stool": { "physics": "passable", "navigation": "obstacle", "semantic_tag": "chair" },
        "chair_garden_iron": { "physics": "passable", "navigation": "obstacle", "semantic_tag": "chair" },
        "table_garden_iron": { "physics": "solid", "navigation": "obstacle", "semantic_tag": "table_garden" },
        "fountain_stone": { "physics": "solid", "navigation": "obstacle", "semantic_tag": "decoration" },
        "lamp_standing": { "physics": "solid", "navigation": "obstacle", "semantic_tag": "light_source" },
        "lamp_street_garden": { "physics": "solid", "navigation": "obstacle", "semantic_tag": "light_source" },
        "door_glass": { "physics": "passable", "navigation": "walkable_door", "semantic_tag": "door_main" },
        "agent_librarian": {"character_name": "Mr. Bookman", "is_agent": True, "soul_file": "librarian_soul.json"},
        "npc_student_f": {"character_name": "Sarah", "is_agent": False, "soul_file": "student_soul.json"},
        "npc_chef": {"character_name": "Gordon", "is_agent": False, "soul_file": "chef_soul.json"},
        "npc_hipster": {"character_name": "Liam", "is_agent": False, "soul_file": "hipster_soul.json"},
    }
}
    \end{lstlisting}
\end{tcolorbox}

\begin{center}
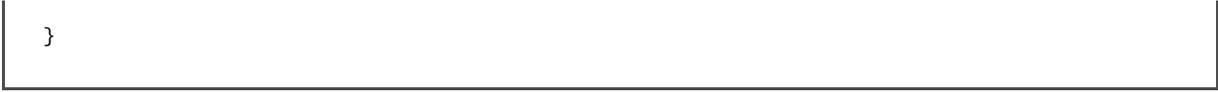

    \captionsetup{hypcap=false}
    \captionof{figure}{Instantiation of the Structured Layout Quadruple .}
    \label{fig:layout}
\end{center}

\clearpage
\twocolumn

\end{document}